\documentclass[12pt,fleqn]{article} 
\usepackage{graphicx} 
\usepackage{amsfonts} 
\usepackage{amssymb} 
\usepackage{epsfig}
\usepackage{cite}

\newcommand{\rsm}{\ensuremath{{\rm SM}}}
\newcommand{\rmssm}{\ensuremath{{\rm MSSM}}}

\newcommand{\tbeta}{\ensuremath{\tan\beta}}
\newcommand{\mSM}{\ensuremath{M_{H_{SM}}}}
\newcommand{\MHpm}{\ensuremath{M_{H^\pm}}}
\newcommand{\ma}{\ensuremath{M_{A^0}}}
\newcommand{\MH}{\ensuremath{M_{H^0}}}
\newcommand{\Mh}{\ensuremath{M_{h^0}}}
\newcommand{\mz}{\ensuremath{M_{Z}}}
\newcommand{\mw}{\ensuremath{M_{W}}}

\def\sw{s_{{\scriptscriptstyle W}}}
\def\cw{c_{{\scriptscriptstyle W}}}
\newcommand{\mhtree}{M^{2 \,{\mathrm{tree}}}_{h^0}}
\newcommand{\mSMtree}{M^{2 \,{\mathrm{tree}}}_{H_{\rsm}}}

\newcommand{\be}{\begin{equation}}
\newcommand{\ee}{\end{equation}}
\newcommand{\bea}{\begin{eqnarray}}
\newcommand{\eea}{\end{eqnarray}}

\begin{document} 
\begin{titlepage} 
\hfill{} 
\begin{tabular}{l} 
KA-TP-25-2001 \\ 
MPI-PhT/2002-29\\
IFT-UAM/CSIC-02-28\,, FTUAM/02-19\\
hep-ph/0208014\\ 
\end{tabular} 
\vspace*{1cm}\\
\begin{center}
\textbf{\large Self-interactions of the lightest $\rmssm$ Higgs boson in the \\
  large pseudoscalar-mass limit}\vspace*{1cm}\\
{\par\centering 
A.~Dobado$^{\,\small{a}}$\,, M.J.~Herrero $^{\,\small{b}}$\,,
W.~Hollik $^{\,\small{c}}$ and S.~Pe{\~n}aranda $^{\,\small{d}}$
\vspace*{0.3cm}~\footnote{electronic addresses:
malcon@fis.ucm.es, herrero@delta.ft.uam.es,
hollik@mppmu.mpg.de, siannah@particle.uni-karlsruhe.de}\\
\par} 
{\par\centering 
$^{\small{a}}$
\textit{Departamento de F{\'\i}sica Te{\'o}rica,\\
Universidad Complutense de Madrid, 28040 Madrid, Spain \vspace*{0.1cm}}}
{\par\centering 
$^{\small{b}}$
\textit{Departamento de F{\'\i}sica Te{\'o}rica,\\
Universidad Aut{\'o}noma de Madrid, Cantoblanco, 28049 Madrid, 
Spain \vspace*{0.1cm}}}
{\par\centering 
$^{\small{c}}$
\textit{Max-Planck-Institut f\"ur Physik,\\
F\"ohringer Ring 6, D-80805 M\"unchen, 
Germany \vspace*{0.1cm}}}
{\par\centering 
$^{\small{d}}$
\textit{Institut f\"{u}r Theoretische Physik, Universit\"{a}t Karlsruhe}\\ 
\textit{Kaiserstra\ss{}e 12, D--76128 Karlsruhe, Germany }}
\end{center}
\vspace*{2cm}
{\par\centering\textbf{\large Abstract}\\ 
\vspace*{0.4cm}
\par} 
\noindent We investigate the decoupling properties of the 
Higgs-sector-induced one-loop corrections 
in the lightest Higgs-boson self-couplings,
in the framework of the Minimal Supersymmetric Standard Model (MSSM). 
The renormalized $n$-point vertex functions with external Higgs particles
in the $\rmssm$ and in the $\rsm$ are derived to the one-loop level and compared in the
$\ma \gg \mz$ limit. The computation has been done in a general $R_{\xi}$ gauge
and the on-shell renormalization scheme is chosen.
By a comparison of the renormalized lightest Higgs-boson $h^{0}$
vertex functions with respect to the
corresponding $\rsm$ ones, we find that the
differences between the predictions of both models are
summarized in the lightest Higgs-boson mass correction $\Delta \Mh$. Consequently, the
radiative corrections are absorbed in the Higgs-boson mass, and
the trilinear and quartic $h^{0}$ self-couplings acquire the same structure
as the couplings of the $\rsm$ Higgs-boson. Therefore, decoupling 
of the heavy $\rmssm$ Higgs bosons occurs and the $\rmssm$ $h^{0}$ 
self-interactions
converge to the $\rsm$ ones in the $\ma \gg \mz$ limit.
\end{titlepage} 

\vfill
\clearpage

\renewcommand{\thefootnote}{\arabic{footnote}}
\setcounter{footnote}{0}

\pagestyle{plain}

\section{Introduction} 

The expectations of the discovery of at least one light Higgs particle at
the next generation of high-energy colliders have greatly increased in recent
years after the valuable data taken at the CERN $e^+ e^-$ Collider LEP and
Fermilab Tevatron~\cite{WGHiggs}. The Standard Model (SM) Higgs-boson mass 
$\mSM$ is now constrained by the worldwide electroweak data to be
$\mSM<195$ GeV and by the direct search performed at the LEP II machine to be 
$\mSM>114.1$ GeV, both at $95\%$ C.L. In the Minimal Supersymmetric Standard
Model (MSSM), on the other hand, the mass of the lightest neutral Higgs
particle, $\Mh$, is predicted to be bounded from above by  $\Mh \lesssim 135$
GeV and the direct searches at LEP give a $95\%$ C.L. exclusion limit of 
$\Mh >91$ GeV. This remarkable shrinkage of the allowed mass range of these
Higgs particles has enhanced even more the expectations for their discovery
at the forthcoming CERN Large Hadron Collider (LHC) 
and the next runs of the Fermilab Tevatron. 

Assuming the hypothetical discovery of one of these two light Higgs particles in
the next generation of colliders, the next challenge will be to measure its 
mass and couplings to all known particles, including its couplings to $\rsm$
fermions and $\rsm$ gauge bosons, as well as the Higgs-particle self-couplings
themselves. The measurement of these parameters can serve to unravel the
supersymmetric (SUSY) or non-supersymmetric origin of this Higgs particle, and, more
specifically, to distinguish if this is the $\rsm$, $H_{\rsm}$, or the $\rmssm$, 
$h^0$. Particularly relevant will be the measurement of the Higgs boson
self-couplings in order to establish the Higgs mechanism experimentally. The
reconstruction of the needed self-interaction potential requires a knowledge
of both the trilinear and quartic self-couplings of the Higgs boson. Since the
predictions of these self-couplings are different in the $\rsm$ and in the
$\rmssm$, their experimental measurement could provide not just an essential way
to determine the mechanism for generating the masses of the fundamental
particles but also an indirect way to test supersymmetry. In the $\rsm$, at
the tree level, the Higgs boson self-couplings are uniquely determined by the
Higgs boson mass $\mSM$ and the vacuum expectation value of the Higgs boson
field $v$, or equivalently the $W$ boson mass $\mw$ and the $SU(2)_L$ gauge
coupling $g$, since $v=2\mw/g$. More specifically, 
$\lambda_{HHH} = 3 \,M^2_{H_{SM}}/v$ and 
$\lambda_{HHHH} =3 \,M^2_{H_{SM}}/v^2$.
In contrast, in the $\rmssm$~\cite{BibliaHiggs}, 
the tree-level trilinear and quartic $h^0$
couplings are determined by the $SU(2)_L$ gauge
coupling $g$, the weak angle $\theta_W$, the $Z$ boson mass $\mz$, the ratio
of the two Higgs-boson vacuum expectation values, $\tan \beta =v_2/v_1$, 
and the CP-odd Higgs-boson mass $\ma$, that is, 
$\lambda_{hhh}^0 = 3 \,({g \mz}/{2 \cw})\,
\cos2\alpha \sin(\beta + \alpha)$ and 
$\lambda_{hhhh}^0= 3 \,({g^2}/{4 \cw^2})\,\cos^2 2\alpha\,$, with the mixing
angle $\alpha$ and $\Mh$ being derived quantities from $\beta\,,\ma$, and $\mz$. 
For arbitrary values of the $\rmssm$ Higgs-sector input parameters $\tan\beta$
and $\ma$, the values of these self-couplings are clearly different from those
of their corresponding trilinear and quartic $\rsm$ couplings. However, the
situation changes in the large pseudoscalar-mass limit $\ma \gg \mz$, 
yielding a particular spectrum with heavy
$H^0\,,H^\pm\,,A^0$ Higgs bosons having similar masses 
$\ma \backsimeq \MHpm \backsimeq \MH$, and a light $h^0$ boson having a
tree-level mass of $\Mh \backsimeq \mz |\cos 2 \beta|$. This  $\ma \gg \mz$
limit is referred to in the literature (and in the present work from now on)
as the {\it decoupling limit} of the MSSM Higgs sector~\cite{dec}, because the 
$h^{0}$ tree-level interactions with the SM fermions and SM gauge bosons
resemble the corresponding SM Higgs boson interactions. Furthermore, 
 in this large pseudoscalar-mass limit, which
also implies $\alpha\rightarrow \beta-\pi/2$, the $h^0$ self-couplings 
approach, respectively,
$\lambda_{hhh}^0 \simeq\,3 g/2\,\mw \,\Mh^2\,,$ and 
$\lambda_{hhhh}^0 \simeq \,3\, g^2/4\,\mw^2 \Mh^2$ and, therefore, they
converge as well to their respective $\rsm$ Higgs boson self-couplings 
if $\mSM$ is taken to be equal to $\Mh$. We can therefore conclude that, 
at the tree level, there is
decoupling of the heavy $\rmssm$ Higgs sector and by studying the light Higgs
boson self-interactions it will be very difficult to unravel its SUSY origin. 

In this paper we are concerned with the behaviour of the self-interactions of
the lightest $\rmssm$ Higgs boson beyond the tree level, where important
radiative corrections from various sectors are expected~\cite{RadCorCouplings,osland,Djouadi,Djouadi2,OtrosH,Selfmt4,Boudjema}. In particular, the
one-loop corrections from the quark and squark sectors are known to be large,
specially in the low $\tan \beta$ and $\ma$ region where they can amount up to
$5\%$ even for heavy squarks in the TeV region~\cite{Selfmt4}. We focus
here on the one-loop radiative corrections to the $h^0$ self-couplings from the
$\rmssm$ Higgs sector itself, and study the decoupling behaviour of these
corrections in the limit where $H^0\,,H^\pm$, and $A^0$ become 
quasi-degenerate and
very heavy as compared to the electroweak scale, while $h^0$ remains light, 
$\Mh \lesssim135$ GeV. We address the question above about the possible
convergence or divergence of these self-couplings to the $\rsm$ ones and 
draw conclusions about the important issue of the possibility of 
discerning between
$h^0$ and $H_{\rsm}$ in the {\it decoupling limit} through the study of their
self-interactions. 

From the more formal point of view of the effective field theory, such study
corresponds to determining the low-energy effective action describing the $h^0$
self-interactions that is obtained after integration of the heavy Higgs-boson
fields, $H^0\,,H^\pm$, and $A^0$, and to concluding if these effective $h^0$
self-interactions, which are valid at low energies $E\ll \ma$, are the same or
not as the $\rsm$ ones. In fact, whenever a
symmetry is present in a fundamental theory and one is interested
in  having this symmetry also in low-energy effective theory,
the particles to be integrated  must satisfy a complete
representation of that symmetry. In our case, the MSSM plays the
role of the fundamental theory and it is $SU(2)_L\times U(1)_Y$ gauge
invariant. Therefore, the SM, which is also gauge invariant, could 
be obtained in principle as an effective theory from the MSSM only
if one integrates both of the Higgs MSSM doublets which include
$H^0 ,\,H^\pm ,\, A^0$, the Goldstone bosons, and the $h^0$ itself, and
not only the heavy modes. This is why we consider here the 
integration of all the MSSM Higgs-boson modes. 

The computation of the low-energy $h^0$
self-interactions can be performed in two ways: either by integrating
out the Higgs-boson fields in the
path integral formalism~\cite{TesisS,Tesis-Proc}, or by 
standard Feynman-diagrammatic methods. We will choose this second method here
and proceed as follows. We evaluate the one-particle irreducible
(1PI) Green functions with external $h^0$ particles to one-loop level
and then we evaluate the corresponding
renormalized 1PI Green functions by fixing the on-shell renormalization
scheme. We will concentrate on studying the behaviour of these 
renormalized vertex functions in the
{\it decoupling limit} where  $H^0\,,H^\pm$ and $A^0$ are much heavier than $Z$,
while both the $h^0$ mass $\Mh$ and the momenta of the external $h^0$
particles remain at some low-energy scale below $\ma$. 
This will give us the values of the low-energy $h^0$ self-couplings that we are
looking for. In order to address the comparison with the $H_{\rsm}$
self-couplings we follow the so-called matching procedure~\cite{Matching} 
in which the quantities to be compared are the renormalized 1PI Green functions
of the two theories. More concretely, we compare here the  
renormalized $h^0$ 1PI Green functions, in the previously mentioned {\it
  decoupling limit}, and the corresponding $\rsm$
renormalized $H_{\rsm}$ 1PI functions at the one-loop level, and we find 
that they are indeed equal for all the studied $n$-point functions $(n=1,...,4)$.
In particular, the $n=3,4$ cases
show  explicitly the convergence  of the
$\rmssm$ $h^0$ self-couplings to the $H_{\rsm}$ self-couplings at the
one-loop level that we are looking for.    
  We also show that all the one-loop effects from the 
heavy Higgs-boson modes $H^0\,,H^\pm$, and $A^0$ in the low-energy $h^0$ 
self-interactions either are absorbed into a redefinition of the low energy
parameters (concretely, $\Mh$), or else are suppressed by inverse powers 
of $\ma$. Consequently, and following the lines of the Appelquist-Carazzone 
theorem~\cite{App-Cara}, we conclude that the   
heavy Higgs bosons $H^0\,,H^\pm,$ and $A^0$ do decouple from the low-energy
$h^0$ self-interactions, not just at the tree level but also at 
one-loop level. This indicates that it will therefore be very difficult, 
even with high-precision
experiments, to distinguish an $h^{0}$ from the
 $\rsm$ Higgs boson by studying their self-interactions, if the pseudoscalar 
 boson mass turns out to be large.

The paper is organized as follows: In section~\ref{sec:Spectrum} we briefly
present those aspects of the $\rmssm$ that we are interested in, fixing our
notation. A discussion of the one-loop $\rmssm$ Higgs-sector contributions 
and the analytical results of these contributions to
the $h^{0}$ self-interactions in the {\it decoupling limit} are included in
subsection~\ref{sec:GreenMSSM}. 
Subsection~\ref{sec:renorMSSM} is devoted to the on-shell
renormalization procedure, where the expressions for the $n$-point vertex
function counterterms, in the $\ma \gg \mz$ limit, and
also the explicit asymptotic results for the 
renormalization constants are presented. Finally, in this subsection
we give the renormalized vertex results in the {\it decoupling limit}.
A discussion of the Higgs-boson self-couplings in the $\rsm$, by giving the
one-loop $H_{\rsm}$ self-couplings corrections, the results for the
renormalization constants by assuming the on-shell scheme, and finally the 
$\rsm$ renormalized vertex functions, is presented in section~\ref{sec:SM}.
A comparison of the results for the renormalized n-point functions in the two
theories is examined and discussed in detail in section~\ref{sec:Matching}.
 Finally, the summary of our conclusions is presented at the end of this last
 section.

\section{$\rmssm$ Higgs sector}
\label{sec:Spectrum}

The Higgs sector of the $\rmssm$
involves two scalar doublets $H_1$ and $H_2$, in order to 
give masses to up- and down-type fermions in a way consistent with supersymmetry. 
The  two-doublet Higgs potential is given by~\cite{BibliaHiggs}
\bea
\label{eq:Higgspot}
V &=& m_1^2 H_1\bar{H}_1 + m_2^2 H_2\bar{H}_2 + m_{12}^2\, (\epsilon_{ab}
      H_1^a H_2^b + {\rm {h.c.}})  \nonumber \\
   && \mbox{} + \frac{g'^2 + g^2}{8}\, (H_1\bar{H}_1 - H_2\bar{H}_2)^2
      +\frac{g^2}{2}\, |H_1\bar{H}_2|^2,
\eea
with the doublet fields $H_1$ and $H_2$,
the soft SUSY-breaking terms $m_1, m_2, m_{12}$, and 
the ${\rm SU(2)_L}$ and ${\rm U(1)_Y}$ gauge couplings $g, g'$.

After spontaneous symmetry breaking 
induced through the neutral
components of $H_1$ and $H_2$ with vacuum expectation values
$v_1$ and $v_2$, respectively, the $\rmssm$ Higgs sector contains five 
physical states: two neutral CP-even scalars
($h^0$ and $H^0$), one CP-odd pseudoscalar ($A^0$), and two charged
Higgs-boson states ($H^{\pm}$). All quartic 
coupling constants are related to the electroweak gauge coupling
constants, thus
imposing various restrictions on the tree-level Higgs-boson masses, 
couplings, and mixing angles.
In particular, all tree-level Higgs-boson parameters can be determined 
in terms of the mass $\ma$ of the CP-odd Higgs boson
[$\ma^2 = m_{12}^2(\tbeta+\cot \beta)$],  and the ratio of the two
Higgs-boson 
vacuum expectation values, $\tbeta = {v_2}/{v_1}$. The other masses and 
the mixing angle $\alpha$ for 
the CP-even states $(h^0,H^0)$ are then fixed, and the trilinear and
quartic self-couplings of the physical Higgs particles can be predicted. 

Our main interest is in the light $h^0$ self couplings,
at the tree-level given by
\be
\label{eq:couptree}
\lambda_{hhh}^0 = 3 \,\frac{g \mz}{2 \cw}
\cos2\alpha \sin(\beta + \alpha)\,\,, \hspace*{1cm}
\lambda_{hhhh}^0= 3 \,\frac{g^2}{4 \cw^2}\cos^2 2\alpha\,.
\ee

In general,
they are different from the tree-level couplings of the $\rsm$ Higgs boson
[see eq.~(\ref{eq:coupSM}) of section~\ref{sec:SM}]. 
However, the situation changes 
in the {\it decoupling limit} of the Higgs sector~\cite{dec}, which implies 
a particular spectrum with very heavy and quasi degenerate
$H^0$, $H^{\pm}$, and $A^0$ Higgs bosons, obeying
\be
\label{eq:heavyM}
\MHpm^2\simeq \ma^2 \left[1+\frac{\mw^2}{\ma^2}\right]\,,\,\,\,\,\,\,
\MH^2\simeq \ma^2 \left[1+(1-\cos^2 2 \beta) \frac{\mz^2}{\ma^2}\right]\,,
\ee
and a light $h^0$ boson, close to the electroweak scale, 
with a tree-level mass of
\be
\label{eq:treemass}
M^{{\mathrm{tree}}}_{h^0}\simeq \mz |\cos2\beta|.
\ee
This limit also implies $\alpha\rightarrow \beta -\pi/2$,
and thus the tree-level self couplings~(\ref{eq:couptree}) tend toward
\be 
\label{eq:treelevelself}
\lambda_{hhh}^0 \simeq
 \,3 \,\frac{g}{2\,\mw} \,\mhtree\,, \quad 
 \lambda_{hhhh}^0 \simeq
\,3\, \frac{g^2}{4\,\mw^2} \mhtree \,.
\ee
Consequently, the tree-level couplings of the light CP-even Higgs boson 
approach the couplings~(\ref{eq:coupSM}) of a $\rsm$ Higgs boson with the same
 mass in the {\it decoupling limit}.

However, there are large radiative corrections contributing to
the  $h^0$ self-couplings. 
The ${\cal O} (m_t^4)$ top-quark and top-squark contributions 
were presented recently in~\cite{Selfmt4}, 
with a discu\-ssion of decoupling of heavy top-squark particles in the one-loop 
contributions. Now we will investigate 
the one-loop contributions to the $h^0$ self-couplings  
originating from the $\rmssm$ Higgs sector itself. Thereby,
in principle, all kinds of diagrams involving gauge bosons, Goldstone bosons,
light and heavy Higgs bosons, have to be taken into account.
Some simplifications can be made, however,
when one studies the deviations of the $\rmssm$ $h^0$ 
self couplings from the corresponding $\rsm$ ones.

(i)
The subset of diagrams with only gauge bosons flowing in the loops and the subset of
diagrams with both gauge and Goldstone bosons give the same
contributions to the $h^{0}$ vertex functions as to the $H_{\rsm}$ vertex
functions, which we have checked by explicit computation.
The only differences come from the extra $\sin(\beta-\alpha)$ 
factors appearing in the $h^{0}$ case, but these factors tend to $1$ 
in the {\it decoupling limit}. Therefore, this kind of diagrams do not 
contribute to the differences between $h^0$  and  $H_{\rsm}$ 
in the {\it decoupling limit}, $\ma \gg \mz$,  and do not need to be 
considered in our analysis. 

(ii)
The contributions from loop diagrams with
heavy Higgs bosons ($H^{0}\,, H^{\pm}\,,A^{0}$) together with gauge bosons 
always go with factors
$\cos (\beta -\alpha)$  and, therefore, they vanish in the large $\ma$
limit. We have also checked this explicitly.
We thus need not consider these diagrams here either.

(iii)
Diagrams involving loops with $\rmssm$ heavy Higgs bosons together with
Goldstone bosons or
the lightest Higgs boson do not appear in the SM.
Contrary to the previous case,
the vertices in these Feynman diagrams are not proportional to
$\cos (\beta -\alpha)$ and they do not vanish in the {\it decoupling limit}. 
These diagrams must therefore be included explicitly in our computation.
Moreover, the purely $\rmssm$ heavy Higgs-boson
one-loop contributions
are obviously an exclusive property of the MSSM 
and thus they have to be taken into account as well.
In addition, there are contributions from 
diagrams involving just Goldstone bosons and the lightest Higgs boson
in the loops. A {\it priori}, they
do not look the same in both models. However,
as we will see in course of the discussion, they converge to
the $\rsm$ ones in the $\ma \gg \mz$ limit (see sections~\ref{sec:GreenMSSM} 
and~\ref{sec:SM}). 

For a transparent discussion, 
we present the details of the computation
in the following subsections.
First, we will give the
one-loop results for the unrenormalized vertex
functions of the lightest Higgs boson $h^0$, by considering the 
limit $\ma \gg \mz$ in the $\rmssm$ Higgs sector.
Then, we give a discussion of the on-shell renormalization scheme
for the $\rmssm$  Higgs bosons 
and list the expressions for the $h^0$  vertex function
counterterms and the explicit results for the renormalization constants
in the large $\ma$ limit (section~\ref{sec:renorMSSM}). Finally, the
renormalized vertex results are given at the end of 
section~\ref{sec:renorMSSM}.

\subsection{Higgs boson self-couplings in the  large $\ma$ limit}
\label{sec:GreenMSSM}

From now on, the general results for the $n$-point ($n=1,...,4$)
renormalized vertex functions
will be summarized by the generic expression
\begin{equation}
\label{eq:notG} 
\Gamma_{\,R\,\,H}^{\,(n)}={\Gamma_{\,0\,\,H}^{(n)}}+\Delta \Gamma_{\,R\,\,H}^{\,(n)}=
{\Gamma_{\,0\,\,H}^{(n)}}+\Delta \Gamma_{H}^{\,(n)}+\delta \Gamma_{H}^{\,(n)}\,,
\end{equation}
where the subscript $R$ denotes renormalized functions, 
the subscript $0$ refers to the tree-level functions, 
the one-loop contributions are summarized 
in $\Delta \Gamma_{H}^{\,(n)}$, and $\delta \Gamma_{H}^{\,(n)}$ represent
the counterterm contributions. The sum of these two last contributions is
denoted by $\Delta \Gamma_{\,R\,\,H}^{\,(n)}$.
Here $H$ refers to the external Higgs-boson particle, 
which corresponds to the lightest CP-even Higgs boson $H\equiv
h^{0}$ in the $\rmssm$ and to the $\rsm$ Higgs boson $H\equiv H_{\rsm}$
in the $\rsm$ case.
The tree-level functions in the $\rmssm$, ${\Gamma_{\,0\,\,h^{0}}^{(n)}}$, 
for $n=3$ and $n=4\,$, are the trilinear and quartic 
$h^{0}$ Higgs self-couplings, already given in~(\ref{eq:couptree}),
and ${\Gamma_{\,0\,h^{0}}^{(2)}}=-(q^2-\mhtree)$. 
Obviously, ${\Gamma_{\,0\,\,H}^{(1)}}=0$. 

\begin{figure}[t]
\begin{center}
\epsfig{file=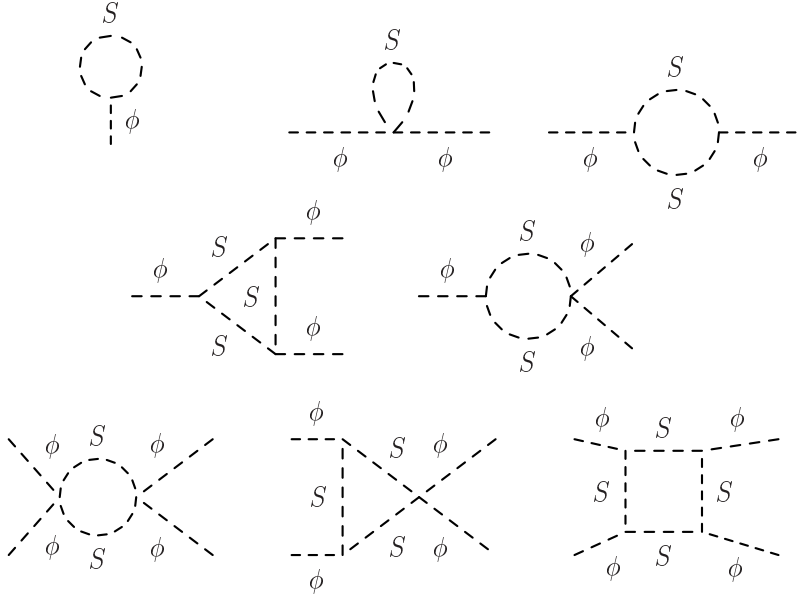,width=12.5cm}
\vspace*{-0.4cm}
\caption{Generic one-loop diagrams contributing to the vertex functions of
the Higgs boson. Here $\phi\equiv h^{0}$ ($\phi\equiv H_{SM}$) in
the $\rmssm$ ($\rsm$) case and correspondingly $S\equiv h^{0}\,, 
G^{0}\,, G^{\pm}\,, H^{0}\,, H^{\pm}\,, A^{0}\,\,
(S\equiv H_{SM}\,,G^{0}\,, G^{\pm})\,$.}
\label{fig:generic}
\end{center}
\end{figure}

We will present in this subsection the results for the 
one-loop contributions $\Delta \Gamma_{h^{0}}^{\,(n)}$  
that come from the diagrams shown generically in Fig.~\ref{fig:generic}.
The computation has been performed by the 
diagrammatic method utilizing {\it FeynArts 3} and {\it FormCalc}~\cite{Hahn}, 
and the results are expressed in terms of the
standard one-loop integrals~\cite{Passarino}. 
We have made the computation in a general $R_{\xi}$ gauge and we have used
dimensional regularization to compute the one-loop integrals. 
Some details of how to compute the integrals in the 
large mass limit $\ma \gg \mz$ can be seen in~\cite{TesisS}.
In the {\it decoupling limit}, 
the heavy Higgs-boson masses have
similar size, up to terms of ${\cal O}({\mz^2}/{\ma^2})$ [see
eq.~(\ref{eq:heavyM})], and correspondingly 
the $\alpha$ angle expansion leads to ${\cal O}({\mz^2}/{\ma^2})$ terms, 
such that
\bea
&&\sin (\beta-\alpha) \simeq 1\,,\,\,\,\,\, 
\cos (\beta-\alpha)\simeq \frac{\mz^2}{\ma^2} S_{2\beta}\,
C_{2\beta}\,,\nonumber\\ 
&&\sin (\beta+\alpha)\simeq - C_{2\beta}\left(1-
\frac{\mz^2}{\ma^2} S_{2\beta}^2\right)\,,\,\,\, 
\cos (\beta+\alpha)\simeq S_{2\beta}\left(1+
\frac{\mz^2}{\ma^2} C_{2\beta}^2\right)\,, \nonumber\\ 
&&\sin 2\alpha  \simeq -S_{2\beta}\left(1+2
\frac{\mz^2}{\ma^2} C_{2\beta}^2\right)\,,\,\,\, 
\cos 2\alpha \simeq - C_{2\beta}\left(1-2
\frac{\mz^2}{\ma^2} S_{2\beta}^2\right)\,.
\eea
Here, and throughout this paper, $C_{2\beta}\equiv \cos 2 \beta$ and 
$S_{2\beta}\equiv \sin 2 \beta$.

Finally, with the explicit results for the one-loop integrals, we obtain 
the contributions to the $n$-point functions.
For a later comparison with the
$\rsm$, it is convenient to split the results according to
\be
\label{eq:notacion}
\Delta \Gamma_{h^{0}}^{(n)}={\Delta \Gamma_{h^{0}}^{(n)}}^{\rm{light}}+
{\Delta \Gamma_{h^{0}}^{(n)}}^{\rm{mixed}}+
{\Delta \Gamma_{h^{0}}^{(n)}}^{\rm{heavy}}\,,
\ee
where ${\Delta \Gamma_{h^{0}}^{(n)}}^{\rm{light}}$ refers to one-loop diagrams
with Goldstone bosons 
($G^{0}, G^{\pm}$) and the lightest Higgs boson $h^0$, 
${\Delta \Gamma_{h^{0}}^{(n)}}^{\rm{mixed}}$ refers to the one-loop 
diagrams involving heavy Higgs particles ($H^0\,,H^\pm\,,A^0$)
together with the $h^{0}$ boson or Goldstone bosons flowing in the loops,
and, finally, ${\Delta \Gamma_{h^{0}}^{(n)}}^{\rm{heavy}}$ refers to the
diagrams with $\rmssm$ purely heavy Higgs contributions (only $H^0\,,H^\pm\,,$ and
$A^0$ in the one-loop diagrams). 

We first list the {\it light} one-loop vertex terms.  Note that for 
our study both the momenta of the external
$h^{0}$ lines and the masses of $h^{0}$, $Z\,, W^\pm$ are quantities to be
considered at the low-energy scales below $\ma$.
The corresponding subset of diagrams is
depicted generically in Fig.~\ref{fig:generic} for the  $\rmssm$ case
by setting $\phi \equiv h^0$ and $S\equiv h^0\,, G^{0}\,, G^{\pm}$. 
The
contributions from the first 3-point diagram and from the last two 4-point
diagrams are UV-finite. The residual diagrams give both a finite contribution and 
a divergent part. Expressed in terms of the scalar
one-loop integrals~\cite{Passarino}
in the convention of~\cite{Denner}, we have, for 
${\Delta \Gamma_{h^{0}}^{(n)}}^{\rm{light}}$  
in the large $M_A$ limit,
\bea
\label{eq:lightMSSM}
&&\hspace*{-2.0cm}{\Delta \Gamma_{h^{0}}^{(1)}}^{\rm{light}}=
\frac{g \mz}{64 \pi^2 \cw}\,C_{2\beta}^2\, \left\{\,
 3 A_{0}(\Mh^2)+A_{0}(\xi \mz^2)+2A_{0}( \xi \mw^2)\,\right\}\,,\nonumber\\
&&\nonumber\\
&&\hspace*{-2.0cm}{\Delta \Gamma_{h^{0}}^{(2)}}^{\rm{light}}=
\frac{g^2}{128 \pi^2 \cw^2}\,C_{2\beta}^2\,\left\{\,
3 A_{0}(\Mh^2)+A_{0}(\xi \mz^2)+2A_{0}( \xi \mw^2)\right.\nonumber\\
&+& \left.\mz^2 C_{2\beta}^2 \left[\, 
9 B_{0}(q^2,\Mh^2,\Mh^2)+B_{0}(q^2,\xi \mz^2,\xi \mz^2)
+2 B_{0}(q^2,\xi \mw^2,\xi \mw^2)\,\right]\right\},\nonumber\\
&&\nonumber\\
&&\hspace*{-2.0cm}{\Delta \Gamma_{h^{0}}^{(3)}}^{\rm{light}}=
\frac{g^3}{256 \pi^2 \cw^3}\mz C_{2\beta}^4 \left\{\,\left[\, 
9 B_{0}(q^2,\Mh^2,\Mh^2)+B_{0}(q^2,\xi \mz^2,\xi \mz^2)
+2 B_{0}(q^2,\xi \mw^2,\xi \mw^2)\,\right.\right.\nonumber\\
&&\hspace*{3.5cm}\left.+(q\rightarrow p)+
(q\rightarrow r)\,\right]\nonumber\\
&+&\left.2 \mz^2 C_{2\beta}^2 \left[\, 
27 C_{0}(q^2,p^2,r^2,\Mh^2,\Mh^2,\Mh^2)+
C_{0}(q^2,p^2,r^2,\xi\mz^2,\xi\mz^2,\xi\mz^2)\right.\right.\nonumber\\
&+&\left.\left.2 C_{0}(q^2,p^2,r^2,\xi\mw^2,\xi\mw^2,\xi\mw^2)\,
\right]\right\},\nonumber\\
&&\nonumber\\
&&\hspace*{-2.0cm}{\Delta \Gamma_{h^{0}}^{(4)}}^{\rm{light}}=
\frac{ g^4}{512 \pi^2 \cw^4}\,C_{2\beta}^4\,\left\{ 
\left[\, 9B_{0}((q+p)^2,\Mh^2,\Mh^2)
+B_{0}((q+p)^2,\xi \mz^2,\xi \mz^2)\right.\right.\nonumber\\
&+&B_{0}((q+p)^2,\xi \mw^2,\xi \mw^2)
+\left.(p \rightarrow r)+(p \rightarrow t)\right]\nonumber\\
&+&2 \mz^2 C_{2\beta}^2  \left[ 
27 C_{0}(q^2,p^2,(q+p)^2,\Mh^2,\Mh^2,\Mh^2)\right.\nonumber\\
&+&C_{0}(q^2,p^2,(q+p)^2,\xi \mz^2,\xi \mz^2,\xi \mz^2)+
2 C_{0}(q^2,p^2,(q+p)^2,\xi \mw^2,\xi \mw^2,\xi \mw^2)\nonumber\\
&&+\left.(p\rightarrow r)+(p\rightarrow t)+
(q\rightarrow t,p\rightarrow r)+(q\rightarrow p,p\rightarrow r)+
 (q\rightarrow p,p\rightarrow t) \right]\nonumber\\
&+&2 \mz^4 C_{2\beta}^4  \left[  
 81D_0(q^2,p^2,r^2,t^2,(q+p)^2,(p+r)^2,
     \Mh^2,\Mh^2,\Mh^2,\Mh^2)\right.\nonumber\\
&+&D_0(q^2,p^2,r^2,t^2,(q+p)^2,(p+r)^2,
     \xi \mz^2,\xi \mz^2,\xi \mz^2,\xi \mz^2)\nonumber\\ 
&+&2D_0(q^2,p^2,r^2,t^2,(q+p)^2,(p+r)^2,
     \xi \mw^2,\xi \mw^2,\xi \mw^2,\xi \mw^2)\nonumber\\
&&+\left.\left.(r\leftrightarrow t)+(p\leftrightarrow r) \right]
\right\}\,.
\eea
Here $q\,,p\,,r\,,$ and $t$ denote the external momenta, 
$\xi$ is the gauge parameter, and $\cw = \cos\theta_W$.
Notice that these contributions are $\xi$-gauge dependent. In addition, 
since they show an explicit dependence on $\beta$, they are different 
from the $\rsm$ ones for arbitrary $\tan\beta$
values. However, as we will see explicitly in 
section~\ref{sec:SM} [see eq.~(\ref{eq:vertexSM})], 
they converge to the $\rsm$ ones in the $\ma \gg \mz$ limit. 
Thus, by identifying the light CP-even Higgs boson mass 
${M^{{\mathrm{tree}}}_{h^0}} \simeq \mz |C_{2\beta}|$ with the 
$\rsm$ Higgs mass $\mSMtree$, the contributions~(\ref{eq:lightMSSM}) 
acquire the structure of the unrenormalized 
$\rsm$ one-loop vertex functions~(\ref{eq:vertexSM}).
Therefore, we conclude that the contributions involving only Goldstone bosons and 
the lightest Higgs boson in the loops
are the same in both models in the $\ma \gg \mz$ limit.
This is equivalent to stating that the
difference between the one-loop unrenormalized $n$-point functions of the
two theories in the {\it decoupling limit} originates only from
diagrams including at least one heavy $\rmssm$ Higgs particle. 
These contributions correspond to 
${\Delta \Gamma_{h^{0}}^{(n)}}^{\rm{heavy}}$ and
${\Delta \Gamma_{h^{0}}^{(n)}}^{\rm{mixed}}$, which read as follows:
\bea
\label{eq:asympresults1p}
&&\hspace*{-1.0cm}{\Delta \Gamma_{h^{0}}^{(1)}}^{\rm{heavy}}=
\frac{g \mz}{32 \pi^2 \cw}\, \left\{ \ma^2 \,
(1+2 \cw^2-3 \,C_{2\beta}^2)\,\left(
\Delta_\epsilon +1-\log\frac{\ma^2}{\mu_{0}^2}\right)\right.\nonumber\\
&&\hspace*{-0.5cm}+\mz^2\left. \left[ 6 \,C_{2\beta}^2 S_{2\beta}^2+
\frac{1}{2}(2-9 \,C_{2\beta}^4+4 \cw^4+
7 \,C_{2\beta}^2-2\,C_{2\beta}^2\, \cw^2)\,
\left(\Delta_\epsilon
  -\log\frac{\ma^2}{\mu_{0}^2}\right)\right]\right\}\,,\nonumber\\
\\
\label{eq:asympresults2p}
&&\hspace*{-1.0cm}{\Delta \Gamma_{h^{0}}^{(2)}}^{\rm{heavy}}=\frac{g^2}{64 \pi^2 \cw^2}\,\left\{
\ma^2 \,(1+2 \cw^2-3 \,C_{2\beta}^2)\,\left(
\Delta_\epsilon +1-\log\frac{\ma^2}{\mu_{0}^2}\right)\right.\nonumber\\
&&\hspace*{-0.5cm}+ \mz^2 \left.\left[ 9 \,C_{2\beta}^2 S_{2\beta}^2+
\frac{1}{2}
(6-3 \,C_{2\beta}^4+12 \cw^4+ \,C_{2\beta}^2-10\,C_{2\beta}^2\, \cw^2)
\left(\Delta_\epsilon-\log\frac{\ma^2}{\mu_{0}^2}\right)\right]\right\}\,,\nonumber\\
&&\hspace*{-1.0cm}{\Delta \Gamma_{h^{0}}^{(2)}}^{\rm{mixed}}=
\frac{g^2}{32 \pi^2 \cw^2}\,\mz^2 \,6\,C_{2\beta}^2 S_{2\beta}^2\,\left(
\Delta_\epsilon +1-\log\frac{\ma^2}{\mu_{0}^2}\right)\,,\\
\nonumber\\
\label{eq:asympresults3p}
&&\hspace*{-1.0cm}{\Delta \Gamma_{h^{0}}^{(3)}}^{\rm{heavy}}=\frac{3 g^3}{64 \pi^2 \cw^3}\,\mz\,
(1+3 \,C_{2\beta}^4 +2 \cw^4-3 \,C_{2\beta}^2-2C_{2\beta}^2 \,\cw^2)
\left(\Delta_\epsilon-\log\frac{\ma^2}{\mu_{0}^2}\right)\,,\nonumber\\
&&\hspace*{-1.0cm}{\Delta \Gamma_{h^{0}}^{(3)}}^{\rm{mixed}}=
\frac{3 g^3}{64 \pi^2 \cw^3}\,\mz \,6\,C_{2\beta}^2 S_{2\beta}^2\,\left(
\Delta_\epsilon +1-\log\frac{\ma^2}{\mu_{0}^2}\right)\,,\\
\nonumber\\
\label{eq:asympresults4p}
&&\hspace*{-1.0cm}{\Delta \Gamma_{h^{0}}^{(4)}}^{\rm{heavy}}=\frac{3 g^4}{128 \pi^2 \cw^4}\,
(1+3 \,C_{2\beta}^4+2 \cw^4-3 \,C_{2\beta}^2-2\,C_{2\beta}^2 \,\cw^2)
\left(\Delta_\epsilon-\log\frac{\ma^2}{\mu_{0}^2}\right)\,,\nonumber\\
&&\hspace*{-1.0cm}{\Delta \Gamma_{h^{0}}^{(4)}}^{\rm{mixed}}=
\frac{3 g^4}{128 \pi^2 \cw^4} \,6\,C_{2\beta}^2 S_{2\beta}^2\,\left(
\Delta_\epsilon +1-\log\frac{\ma^2}{\mu_{0}^2}\right)\,.
\eea

Obviously, ${\Delta \Gamma_{h^{0}}^{(1)}}^{\rm{mixed}}=0$. 
Here $\mu_{0}$ denotes the scale of dimensional regularization and the singular 
$\Delta_\epsilon$ term is defined, as usual, by
\begin{equation} 
\label{eq:Delta}
\hspace*{0.6cm}  
\displaystyle {\Delta}_\epsilon=\frac{2}{\epsilon }-{\gamma }_{\epsilon}  
+\log (4\pi)\,, \hspace*{0.3cm} \epsilon = 4-D\,. 
\end{equation}

Terms that are suppressed by inverse powers of the heavy mass $\ma$
and thus vanish in the {\it decoupling limit} 
are dropped in the expressions given above.
Contrary to the one-loop contributions from diagrams with Goldstone bosons and 
the lightest $h^{0}$ Higgs particle~(\ref{eq:lightMSSM}),
the above $h^{0}$ vertex function contributions are 
$\xi$-gauge independent. The Feynman diagrams contributing to
the one-loop $\rmssm$ heavy Higgs-boson sector part, 
${\Delta \Gamma_{h^{0}}^{(n)}}^{\rm{heavy}}\, (n=1,...,4)$, 
appearing in eqs.~(\ref{eq:asympresults1p})-(\ref{eq:asympresults4p}) can be
extracted from Fig.~\ref{fig:generic} by choosing $\phi\equiv h^{0}$ and
$S\equiv H^{0}\,, H^{\pm}\,, A^{0}$.
The contributions from the first diagram in the 3-point function and from the
last two diagrams in the 4-point function are finite and vanish in the 
$\ma \gg \mz$ limit. 
The remaining diagrams are UV-divergent and contain  
a logarithmic dependence on the heavy pseudoscalar mass $\ma$ and,
for $n=1,2$, 
a quadratic dependence on $\ma$ as well. In contrast, the {\it mixed} diagrams 
do not give $\ma^2$ terms, but they are logarithmically dependent on $\ma$.
The corresponding specific Feynman diagrams are obtained from Fig.~\ref{fig:generic}
by taking $\phi\equiv h^{0}$ and accordingly to the {\it light} and {\it heavy}
particles that can be flowing in the loops,
$S\equiv h^{0}\,,G^{0}\,,G^{\pm}\,,H^{0}\,, H^{\pm}\,, A^{0}$.
More specifically, the {\it mixed} diagrams that give contributions 
different from zero in the {\it decoupling limit} correspond to the 
third, fifth, and sixth diagrams in Fig.~\ref{fig:generic}, with $h^{0}$ and $H^{0}$
($S\equiv h^{0}\,,H^{0}$), $G^{0}$ and $A^{0}$
($S\equiv G^{0}\,,A^{0}$), and $G^{\pm}$ and $H^{\pm}$
($S\equiv G^{\pm}\,,H^{\pm}$), in the two internal propagators of the loops.

Let us remark that, in these results for the unrenormalized vertex functions,  
all the potential
non-decoupling effects of the heavy Higgs $\rmssm$ particles  manifest 
as some divergent contributions in $D=4$ and some finite contributions,
one of which is 
logarithmically dependent on the heavy pseudoscalar Higgs-boson mass $\ma$ and the
other one is quadrati\-cally dependent on $\ma$.
Obviously, all the results displayed up to now are, in general, UV-divergent.
In order to get finite 1PI Green functions and finite predictions for 
physical observables, 
renormalization has to be performed by adding appropriate counterterms. 
This is the subject of the next subsection.

\subsection{Renormalization in the $\rmssm$.} 
\label{sec:renorMSSM}

For a systematic one-loop computation, the free parameters of the Higgs
potential  $m_1^2,\, m_2^2,\, m_{12}^2,$ $\, g,\, g'$
and the two vacua $v_1,\, v_2$ are replaced by the corresponding 
renormalized parameters
plus counterterms. This transforms the potential $V$ into $V +\delta V$, 
where $V$ is expressed in terms of the renormalized parameters,
and $\delta V$ is the counterterm potential. 
By using the standard renormalization 
procedure~\cite{CPR,Dabelstein} 
\bea
&&m_i^2 \rightarrow Z_{H_{i}}^{-1} (m_i^2+\delta m_i^2)\,,\,\,\,\,
m_{12}^2 \rightarrow Z_{H_{1}}^{-1/2} Z_{H_{2}}^{-1/2}
(m_{12}^2+\delta m_{12}^2)\,,\nonumber\\
&&v_i \rightarrow Z_{H_{i}}^{1/2}(v_i -\delta v_i)\,,\,\,\,\,
g \rightarrow Z_{1}^{W} Z_{2}^{W\, -3/2} g\,,\,\,\,\,
g' \rightarrow Z_{1}^{B} Z_{2}^{B\,-3/2} g'\,,
\eea
with Higgs-field renormalization constants $\delta Z_{H_{i}}$, and
by using the minimum condition on the potential at tree level, 
we obtain the counterterms for the $n$-point $(n=1,...,4)$ vertex
functions.
The results in the {\it decoupling limit} are:
\bea
\label{eq:counter}
\delta \Gamma_{h^{0}}^{(1)}&=&
\frac{g\mz}{2\cw} \,C_{2\beta}\,
v^2 \left(\sin^2\beta \,\delta Z_{H_{2}}-
\cos^2\beta \,\delta Z_{H_{1}}\right)-v \,\delta M_{12}^2\nonumber\\
&+&\frac{1}{4} \frac{g^2}{\cw^2} v^2 \,C_{2\beta}^2\, \delta v
- \frac{1}{8} \,v^3 \,C_{2\beta}^2\, \delta G^2\nonumber\\
&+&C_{2\beta}\,S_{2\beta}\,\frac{\mz^2}{\ma^2}\,\left[
\frac{g^2}{16\cw^2} S_{2\beta}\,v^3\,\left(\delta Z_{H_{2}} (2 C_{2\beta}-1)
+\delta Z_{H_{1}} (2 C_{2\beta}+1)\right)\right.\nonumber\\
&-& \left.v \,\delta C_{12}^2-
\frac{g^2}{4\,\cw^2} \,C_{2\beta}\,S_{2\beta}\,v^2\, \delta v+
\frac{1}{8} \,v^3 \,C_{2\beta}\,S_{2\beta}\, \delta G^2\right]\,,\nonumber\\
\delta \Gamma_{h^{0}}^{(2)}&=&
q^2 \,\left[\,\left(\sin^2\beta \,\delta Z_{H_{2}}+
\cos^2\beta \,\delta Z_{H_{1}}\right)
+ \frac{\mz^2}{\ma^2}\,C_{2\beta}\,S_{2\beta}^2\,
\left(\delta Z_{H_{2}}-\delta Z_{H_{1}}\right)\right]\nonumber\\
&+&\frac{3}{4} \left[\,C_{2\beta}\,
v^2\,\frac{g^2}{\cw^2}\left(\sin^2\beta \,\delta Z_{H_{2}}-
\cos^2\beta \,\delta Z_{H_{1}}\right)-
\frac{4}{3}\,\delta M_{12}^2 \right.\nonumber\\
&+&\left. \frac{g^2}{\cw^2} \,C_{2\beta}^2 \,v\,\delta v
- \frac{v^2}{2} \,C_{2\beta}^2 \,
\delta G^2\right]\nonumber\\
&+&C_{2\beta}\,S_{2\beta}\,\frac{\mz^2}{\ma^2}\,\left[
\frac{3 g^2}{8\cw^2} S_{2\beta}\,v^2\,\left(\delta Z_{H_{2}} (2 C_{2\beta}-1)
+\delta Z_{H_{1}} (2 C_{2\beta}+1)\right)\right.\nonumber\\
&-& \left.2\,\delta C_{12}^2-
\frac{3 g^2}{2\,\cw^2} \,C_{2\beta}\,S_{2\beta}\,v\, \delta v+
\frac{3}{4} \,v^2 \,C_{2\beta}\,S_{2\beta}\, \delta G^2\right]\,,\nonumber\\
\delta \Gamma_{h^{0}}^{(3)}&=& \frac{3}{4} \,C_{2\beta} \, \left[
2v \frac{g^2}{\cw^2}\left(\sin^2\beta \,\delta Z_{H_{2}}-
\cos^2\beta \,\delta Z_{H_{1}}\right)
+\frac{g^2}{\cw^2} \,C_{2\beta} \, \delta v- v\,C_{2\beta}\,
\delta G^2\right]\nonumber\\
&+& \frac{9}{4} \,C_{2\beta} \,S_{2\beta}\,
\frac{\mz^2}{\ma^2}\,\left[
\frac{g^2}{4\cw^2} S_{2\beta}\,v\,\left(\delta Z_{H_{2}} (2 C_{2\beta}-1)
+\delta Z_{H_{1}} (2 C_{2\beta}+1)\right)\right.\nonumber\\
&-& \left.
\frac{g^2}{\cw^2} C_{2\beta}\,S_{2\beta} \delta v+
v \,C_{2\beta}\,S_{2\beta}\, \delta G^2\right]\,,\nonumber\\
\delta \Gamma_{h^{0}}^{(4)}&=& \frac{3}{4} \,C_{2\beta}\,\left[
2 \frac{g^2}{\cw^2}\left(\sin^2\beta \,\delta Z_{H_{2}}-
\,\cos^2\beta \,\delta Z_{H_{1}}\right)-\,C_{2\beta} \,\delta
G^2\right]\nonumber\\
&+& \frac{9}{4} \,C_{2\beta} \,S_{2\beta}\,
\frac{\mz^2}{\ma^2}\,\left[
\frac{g^2}{4\cw^2} S_{2\beta}\,\left(\delta Z_{H_{2}} (2 C_{2\beta}-1)
+\delta Z_{H_{1}} (2 C_{2\beta}+1)\right)
+ C_{2\beta}\,S_{2\beta}\, \delta G^2\right].\nonumber\\
\eea
Here all ${\cal O}(\mz^2/\ma^2)$ contributions are still
explicitly included. We have introduced the abbreviations
\bea
\label{eq:abrev}
\delta G^2 &\equiv& \delta g^2+\delta g'^2 =
g^2 (2 \,\delta Z_{1}^{W}-3 \,\delta Z_{2}^{W})- 
g'^2 \delta Z_{2}^{B}\,,\nonumber\\
\delta M_{12}^2 &\equiv& \cos^2 \beta \,\delta m_1^2 +
\sin^2 \beta \,\delta m_2^2+S_{2\beta} \,\delta m_{12}^2\,,\nonumber\\
v \,\delta v &=& v_1 \delta v_1 + v_2 \delta v_2 \,\,\,\,{\mbox{with}}\,\,\,\,
v^2=v_1^2+v_2^2\,,\nonumber\\
\delta C_{12}^2 &\equiv& C_{2\beta} \,\delta m_{12}^2+
\frac{S_{2\beta}}{2} (\delta m_2^2-\delta m_1^2)\,.
\eea

Correspondingly, the tadpole counterterm for the 
$H^0$ Higgs boson, $\delta {\Gamma_{H^{0}}^{(1)}}$, and 
the counterterm for the pseudoscalar two-point function, 
$\delta {\Gamma_{A^{0}}^{(2)}}$, which are necessary for the $\rmssm$ on-shell
renormalization, are given, in the {\it decoupling limit}, by
\bea
\label{eq:tapH0}
\delta \Gamma_{H^{0}}^{(1)}&=&
-\frac{g\mz}{8\cw} \,S_{2\beta}\,
v^2 \left[\, \left(-1+ C_{2\beta}\right) \,\delta Z_{H_{2}}+
\left(1+ C_{2\beta}\right)\,\delta Z_{H_{1}}\,\right]\nonumber\\
&+&v \,\delta C_{12}^2
+\frac{1}{4} \frac{g^2}{\cw^2} v^2 \,C_{2\beta}\,S_{2\beta}\, \delta v
- \frac{1}{8} \,v^3 \,C_{2\beta}\,S_{2\beta}\, \delta G^2\nonumber\\
&-&C_{2\beta}\,S_{2\beta}\,\frac{\mz^2}{\ma^2}\,\left[
\frac{g^2}{4\cw^2} (1-2 C_{2\beta}^2)\,v^3\,
\left(\sin^2 \beta\,\delta Z_{H_{2}}
-\cos^2 \beta\,\delta Z_{H_{1}}\right)\right.\nonumber\\
&+& \left.v \,\delta M_{12}^2-
\frac{g^2}{4\,\cw^2} \,C_{2\beta}^2\,v^2\, \delta v+
\frac{1}{8} \,v^3 \,C_{2\beta}^2\, \delta G^2\right]\,,\\
\label{eq:counmA2p}
\delta {\Gamma_{A^{0}}^{(2)}}&=&
q^2 \,\left(\sin^2\beta \,\delta Z_{H_{1}}+
\cos^2\beta \,\delta Z_{H_{2}}\right)
-\frac{1}{2}\left(\sin^2 \beta \,\delta m_1^2+\cos^2 \beta \,\delta m_2^2-
\sin 2 \beta \,\delta m_{12}^2 \right)\nonumber\\
&+&\frac{1}{8} \frac{g^2}{\cw^2} v^2 C_{2\beta}^2 \left(
\frac{\cw^2}{g^2} \,\delta G^2 +\delta Z_{H_{1}}+\delta Z_{H_{2}}
-2 \frac{\delta v}{v}\right)\,.
\eea

We note that no ${\cal O}(\ma^2)$ contributions 
to the renormalization constants $\delta G^2,\, \delta v$, and
$\delta Z_{H_{i}}\,(i=1,2)$ exist. Therefore, terms of the type
${\cal O}(\frac{\mz^2}{\ma^2}) \, \cdot\delta G^2$,
${\cal O}(\frac{\mz^2}{\ma^2}) \,\cdot \delta v$ or
${\cal O}(\frac{\mz^2}{\ma^2}) \, \cdot \delta Z_{H_{i}}$ 
in~(\ref{eq:counter}) and~(\ref{eq:tapH0}) can be safely neglected. 

In the on-shell scheme, the counterterms are fixed by imposing the
following renormalization conditions~\cite{Dabelstein,Renor}:\\
-- the on-shell conditions for $M_{W,Z}$ and the electric charge $e$, as
in the $\rsm$;\\ 
-- the on-shell condition for the $A^0$ boson with the pole mass $M_A$;\\
-- the tadpole conditions for vanishing renormalized tadpoles for both the $H^0$
and $h^0$ Higgs fields, i.e.\
the sum of the one-loop tadpole diagrams for $H^0$,
$h^0$ and the corresponding tadpole counterterms is equal to zero; and,\\
-- the renormalization of $\tan\beta $ in such a way 
that the relation $\tan\beta= v_2/v_1$ is valid for the true one-loop 
Higgs minima.

Notice that the above condition for vanishing renormalized tadpole
diagrams ensures that $v_1, v_2$ determine the minimum of the one-loop potential. The
relation $\tan\beta= v_2/v_1$ in terms of the ``true vacua'' is maintained by
the condition $\delta v_1/v_1 =\delta v_2/v_2$.
By the above set of conditions, the input for the
$\rmssm$ Higgs sector is fixed by the $A^0$ pole mass $M_A$ and $\tan\beta$, together
with the standard gauge-sector input $M_{W,Z}$ and $e$. 

In order to compute the renormalization constants
$\delta Z_{H_{1}}\,, \delta Z_{H_{2}}\,,
\delta G^2$, and $\delta v$, we express them in terms of 
the vector boson self-energies, the $A^0$-boson self-energy, and 
the $A^0 Z$ non-diagonal self-energy~\cite{Dabelstein}:
\bea
\label{eq:countoneshell}
\delta Z_{H_{1}}&=& -\Sigma_{A^{0}}^{'}(\ma^2)-\frac{\cot \beta}{\mz}
\,\Sigma_{A^{0}Z} (\ma^2)\,,\nonumber\\
\delta Z_{H_{2}}&=& -\Sigma_{A^{0}}^{'}(\ma^2)+\frac{\tan \beta}{\mz}
\,\Sigma_{A^{0}Z} (\ma^2)\,,\nonumber\\
\delta G^2&=&\frac{g^2}{\cw^2}\left[
\Sigma_{\gamma}^{'}(0)-2\frac{\sw}{\cw}
\frac{\Sigma_{\gamma Z}(0)}{\mz^2}-
\frac{\cw^2-\sw^2}{\sw^2}\left(\frac{\Sigma_{Z}(\mz^2)}{\mz^2}-
\frac{\Sigma_{W}(\mw^2)}{\mw^2}\right)\right]\,,\nonumber\\
2\frac{\delta v}{v}&=&-\Sigma_{A^{0}}^{'}(\ma^2)+
\frac{\tan \beta-\cot \beta}{\mz}\,\Sigma_{A^{0}Z} (\ma^2)\nonumber\\
&+&\Sigma_{\gamma}^{'}(0)-2\frac{\sw}{\cw}
\frac{\Sigma_{\gamma Z}(0)}{\mz^2}-\frac{\cw^2}{\sw^2}
\frac{\Sigma_{Z}(\mz^2)}{\mz^2}
+\frac{\cw^2-\sw^2}{\sw^2}\frac{\Sigma_{W}(\mw^2)}{\mw^2}\,.
\eea

Partial results for the one-loop contributions to the vector boson self-energies
can be extracted from the last article in Ref.~\cite{TesisS} or from
the first article in Ref.~\cite{Dabelstein}. We have recalculated explicitly  
all the self-energies that appear in~(\ref{eq:countoneshell}), and 
we have checked that our results agree with previous ones in the literature.
Here we do not present the intermediate results, but list 
only the final expressions for the  counterterms.

First, we found that $\delta Z_{H_{1,2}}$ get contributions that are 
suppressed by inverse powers of the heavy mass $\ma$. 
However,
such terms 
of order ${\cal O}\left({\mz^2}/{\ma^2}\right)$ to $\delta Z_{H_{1,2}}$
are relevant in order to implement consistently the $A^{0}$-boson on-shell condition
$\Delta \Gamma_{A^{0}}^{(2)}(\ma^2)+\delta {\Gamma_{A^{0}}^{(2)}}(\ma^2)=0$. 
Their expressions are given explicitly in the Appendix A.

The various contributions to $\delta v$ and $\delta G^2$
are split again into ``light'' and ``heavy''. 
The light ones originate from diagrams involving  Gold\-stone bosons 
and the lightest $h^{0}$ Higgs boson in the loops,
\bea
\label{eq:vyGMSSM}
\hspace*{-1.0cm}\frac{\delta v}{v}^{\rm{light}}&=& 
-\frac{g^2}{128 \cw^2\sw^2\pi^2}\frac{1}{\mz^2}
\left\{\sw^2\left[A_{0}(\Mh^2)+A_{0}(\xi\mz^2)\right]\right.\nonumber\\
&&-2(1-\cw^2-4\cw^4+4\cw^6)\,A_{0}(\xi\mw^2)+
\frac{4}{3}\mz^2\cw^2\sw^4\,B_{0}(0,\xi\mw^2,\xi\mw^2)\nonumber\\
&&+4\cw^2\left[(1-2\cw^2)^2\,B_{22}(\mz^2,\xi\mw^2,\xi\mw^2)
+B_{22}(\mz^2,\Mh^2,\xi\mz^2)\right.\nonumber\\
&&\left.\left.+(1-2\cw^2)\left(B_{22}(\mw^2,\Mh^2,\xi\mw^2)+
B_{22}(\mw^2,\xi\mz^2,\xi\mw^2)\right)\right]\right\},\\
\nonumber\\
\hspace*{-1.0cm}\delta G^{2\,\rm{light}}&=& \frac{g^4}{16 \cw^4 \sw^2\pi^2}
\frac{1}{\mz^2}\left\{2\cw^2(1-3\cw^2+2\cw^4)\,A_{0}(\xi\mw^2)\right.\nonumber\\
&&-\frac{1}{3}\cw^2\sw^4\mz^2\,B_{0}(0,\xi\mw^2,\xi\mw^2)\nonumber\\
&&-(1-2\cw^2)\left[B_{22}(\mw^2,\Mh^2,\xi\mw^2)-
B_{22}(\mz^2,\Mh^2,\xi\mz^2)\right.\nonumber\\
&&+\left.\left.B_{22}(\mw^2,\xi\mz^2,\xi\mw^2)-(1-2\cw^2)^2\,
B_{22}(\mz^2,\xi\mw^2,\xi\mw^2)\right]\right\}\,.
\eea

``Mixed'' contributions from
diagrams that contain a heavy Higgs-boson together with a Goldstone boson or
the light $h^{0}$ in the loops do not contribute to either of these two
renormalization constants. Such diagrams are suppressed by the factor
$\cos (\beta-\alpha)$ and therefore they vanish in the 
{\it decoupling limit}.
Purely heavy Higgs-boson contributions to 
$\delta v$ are of order ${\cal O}\left({\mz^2}/{\ma^2}\right)$ 
in the $\ma \gg \mz$ limit, and therefore they also vanish. 
In contrast, for $\delta G^2$ we get a non-vanishing contribution,
\be
\label{eq:asympcount}
 \delta G^{2\,\rm{heavy}} = \frac{g^4}{96 \pi^2 \cw^4}\,
(1+2\cw^4-2\cw^2)\,\left(
\Delta_\epsilon -\log\frac{\ma^2}{\mu_{0}^2}\right)\,.
\ee 

The only remaining parameters in~(\ref{eq:counter})
still to be fixed are the mass counterterms $\delta m_1^2\,,\delta m_2^2,$ and 
$\delta m_{12}^2$. Their expressions are derived from the
conditions for $H^0$ and $h^{0}$ vanishing renormalized tadpole diagrams
and from the
on-shell condition for the $A^0$ boson. 
The explicit results for these mass counterterms are given 
in~(\ref{eq:mslight})-(\ref{eq:msheavy}) of the Appendix A.
For completeness, the $H^0$ tadpole and the 
$A^0$ self-energy one-loop results are also presented at the beginning 
of the Appendix A. Then, by implementing all the
renormalization constants in~(\ref{eq:counter}), we obtain
the vertex function counterterms, separated into {\it light},
{\it mixed} and {\it heavy} contributions in the $\ma \gg \mz$ limit.
The one-point counterterm has already been used  
for the determination of the basic renormalization constants and is not required 
for the further discussion; we thus do not list it here:
\bea
\label{eq:redeflight}
\delta {\Gamma_{h^{0}}^{\,(2)}}^{\rm{light}}&=&
-\frac{g^2}{128 \pi^2 \cw^2}\,C_{2\beta}^2\,\left\{\,
A_{0}(\Mh^2)-A_{0}(\xi \mz^2)-(2-16\cw^2+16\cw^4)A_{0}( \xi \mw^2)\right.\nonumber\\
&& \left.+8 B_{22}(\mz^2, \Mh^2,\xi \mz^2)+
8(1-2\cw^2)^2 B_{22}(\mz^2, \xi \mw^2,\xi \mw^2)\,\right\},\nonumber\\
 \nonumber\\
\delta {\Gamma_{h^{0}}^{\,(3)}}^{\rm{light}}&=& 
\frac{3g^3}{256 \pi^2 \cw^3}\,\frac{1}{\mz}\,C_{2\beta}^2\left\{\,
A_{0}(\Mh^2)+A_{0}(\xi \mz^2)\right.\nonumber\\
&&+\frac{1}{\sw^2}(2-18\cw^2+40\cw^4-24\cw^6)A_{0}( \xi \mw^2)\nonumber\\
&&+\frac{4}{3}\cw^2\sw^2\mz^2B_{0}(0,\xi \mw^2,\xi \mw^2)
+4\frac{1}{\sw^2}(3\cw^2-2)B_{22}(\mz^2,\Mh^2,\xi \mz^2)\nonumber\\
&&+4(1-2\cw^2)\frac{1}{\sw^2}\left[B_{22}(\mw^2,\Mh^2,\xi \mw^2)+
B_{22}(\mw^2, \xi \mz^2,\xi \mw^2)\right.\nonumber\\
&&+\left.\left.(1-2\cw^2)(3\cw^2-2)B_{22}(\mz^2, \xi \mw^2,\xi \mw^2)\right]\right\}\,,
\nonumber\\\nonumber\\
\delta {\Gamma_{h^{0}}^{\,(4)}}^{\rm{light}}&=& 
\frac{3g^4}{256\pi^2\cw^4 }\,\frac{1}{\mz^2}\,C_{2\beta}^2\left\{\,
-8\frac{\cw^2}{\sw^2}(1-3\cw^2+2\cw^4)A_{0}( \xi \mw^2)\right.\nonumber\\
&&+\frac{4}{3}\cw^2\sw^2\mz^2B_{0}(0,\xi \mw^2,\xi \mw^2)\nonumber\\
&&+4(1-2\cw^2)\frac{1}{\sw^2}\left[B_{22}(\mw^2,\Mh^2,\xi \mw^2)+
B_{22}(\mw^2, \xi \mz^2,\xi \mw^2)\right.\nonumber\\
&&\left.\left.-B_{22}(\mz^2,\Mh^2,\xi \mz^2)-(1-2\cw^2)^2
B_{22}(\mz^2, \xi \mw^2,\xi \mw^2)\right]\right\}\,,
\eea

\be
\label{eq:redef234pmixed}
\delta {\Gamma_{h^{0}}^{\,(2)}}^{\rm{mixed}}=0\,\,,\,\,\,\,
\delta {\Gamma_{h^{0}}^{\,(3)}}^{\rm{mixed}}=0\,\,,\,\,\,\,
\delta {\Gamma_{h^{0}}^{\,(4)}}^{\rm{mixed}}=0\,\,,
\ee
\bea
\label{eq:redef234pheavy}
\delta {\Gamma_{h^{0}}^{\,(2)}}^{\rm{heavy}}&=&
-\frac{g^2}{64 \pi^2 \cw^2}\,\left\{
\ma^2 \,(1+2 \cw^2-3 \,C_{2\beta}^2)\,\left(
\Delta_\epsilon +1-\log\frac{\ma^2}{\mu_{0}^2}\right)\right.\nonumber\\
&&+ \mz^2 \left[ 9 \,C_{2\beta}^2 S_{2\beta}^2+
\frac{1}{6}\left(6-45 \,C_{2\beta}^4+12 \cw^4+ 43\,C_{2\beta}^2-14\,
C_{2\beta}^2\, \cw^2\right.\right.\nonumber\\
&&+\left.\left.8\,C_{2\beta}^2\, \cw^4)
\left(\Delta_\epsilon-\log\frac{\ma^2}{\mu_{0}^2}\right)\right]\right\}\,,\nonumber\\
\delta {\Gamma_{h^{0}}^{\,(3)}}^{\rm{heavy}}&=&
-\frac{g^3}{64 \pi^2 \cw^3}\,\mz \,C_{2\beta}^2\,(1+2\cw^4-2\cw^2)\,\left(
\Delta_\epsilon -\log\frac{\ma^2}{\mu_{0}^2}\right)\,,\nonumber\\
\delta {\Gamma_{h^{0}}^{\,(4)}}^{\rm{heavy}}&=&
-\frac{g^4}{128 \pi^2 \cw^4}\,C_{2\beta}^2\,(1+2\cw^4-2\cw^2)\,\left(
\Delta_\epsilon -\log\frac{\ma^2}{\mu_{0}^2}\right)\,.
\eea
The {\it heavy} contributions contain, in addition to the 
singular $\Delta_\epsilon$ part, finite logarithmic heavy mass
terms, and for the two-point function quadratic heavy-mass terms also. 

The renormalized vertex functions in the $\ma \gg \mz$ limit can now 
be obtained simply by adding the one-loop 
contributions~(\ref{eq:lightMSSM})-(\ref{eq:asympresults4p})
and the counterterms~(\ref{eq:redeflight})-(\ref{eq:redef234pheavy}).
Since it is just an algebraic substitution, we do not present these
results explicitly here. However, some comments are in order. First,
the quadratic heavy mass terms, ${\cal{O}}(\ma^2)$, in the two-point
result cancel  once we add the one-loop result
in~(\ref{eq:asympresults2p}) and the counterterm~(\ref{eq:redef234pheavy}). 
Thus, there are no ${\cal O}(\ma^2)$ terms
left in any of the renormalized $n$-point functions in the 
large $\ma$ limit, but still the logarithmic dependence on $\ma$ remains.
Second, the
renormalized $h_{0}$ Higgs-boson self-energy evaluated at the physical mass 
$\Mh$ allows us
to define the $\rmssm$ Higgs-boson mass correction $\Delta \Mh^2$ such that 
$\Mh^2=\mhtree+\Delta \Mh^2$ and 
$\Delta \Mh^2=\Delta \Gamma_{\,R\,\,h^{0}}^{\,(2)}(\Mh^2)$. 
Evaluating the renormalized $h^{0}$
two-point function at $q^2=\Mh^2$, we get the following
one-loop mass correction for the light Higgs boson:
\be
\label{eq:firstdefMh}
\Delta \Mh^2= \frac{g^2}{32 \pi^2 \cw^2}\,\mz^2\,\left(\,\Psi_{\rm{light}}+
\Psi_{\rm{mixed}}+\Psi_{\rm{heavy}}\right)\,,
\ee 
with
\bea
\label{eq:contrib-redef}
\hspace*{-1.0cm}\Psi_{\rm{light}}&=&
\frac{1}{4\mz^2}\,C_{2\beta}^2\,\left\{\,
2 A_{0}(\Mh^2)+4(1-2\cw^2)^2A_{0}( \xi \mw^2)
+2A_{0}( \xi \mz^2)\right.\nonumber\\
&-& 8 B_{22}(\mz^2, \Mh^2,\xi \mz^2)-
8(1-2\cw^2)^2 B_{22}(\mz^2, \xi \mw^2,\xi \mw^2)\nonumber\\
&+&\left. \mz^2C_{2\beta}^2 \left[ 9B_{0}(\Mh^2,\Mh^2,\Mh^2)+B_{0}(\Mh^2,\xi \mz^2,\xi \mz^2)
+2 B_{0}(\Mh^2,\xi \mw^2,\xi \mw^2)\,\right]\,\right\},\nonumber\\
\hspace*{-1.0cm}\Psi_{\rm{mixed}}&=&6\, C_{2\beta}^2\, S_{2\beta}^2 
\left(\Delta_\epsilon+1-\log\frac{\ma^2}{\mu_{0}^2}\right)\,,\nonumber\\
\hspace*{-1.0cm}\Psi_{\rm{heavy}}&=&
\left(1+3\,C_{2\beta}^4 +2 \cw^4-\frac{10}{3} \,C_{2\beta}^2-
\frac{4}{3} \,C_{2\beta}^2\, \cw^2-
\frac{2}{3} \,C_{2\beta}^2\, \cw^4\right)
\left(\Delta_\epsilon-\log\frac{\ma^2}{\mu_{0}^2}\right)\,.
\eea

This mass correction  is still UV-divergent since  
we have not included the complete set of diagrams, restricting ourselves
to the subset providing contributions that are different from those in 
the $\rsm$. 
We have checked explicitly that for cancellation of the divergences 
in the renormalized two-point function 
it is necessary to include the subset of 
one-loop diagrams accounting for the gauge-boson contributions. 
We have also checked that when the gauge-boson contributions are included the
$\xi-$gauge dependence in the {\it light} one-loop renormalized $2-$point
function disappears. By including all 1PI one-loop contributions, we have
checked as well that our results are in agreement with the complete results for
the radiative corrections to the Higgs-boson mass listed in the 
literature~\cite{Higgsoneloop,Heinemeyer}.

On the other hand,
the contributions from
one-loop diagrams with have at least one heavy Higgs particle 
($\Psi_{\rm{mixed}}$ and $\Psi_{\rm{heavy}}$) contain, in addition to the 
singular $\Delta_\epsilon$ term, some logarithmic heavy mass
terms that appear like as non-decoupling effects 
of the heavy particles at the renormalized level of renormalized Green functions. 
These apparently 
non-decoupling effects, however, are not physically observable since they are 
absorbed into redefinitions of the low energy parameters, more
specifically, in the redefinition of the $h^0$ mass,
 \be\label{eq:replaceMh0}\Mh^2 ={\Mh^2}^{\rm tree}+\Delta \Mh^2\,,\ee
with $\Delta \Mh^2$ given in~(\ref{eq:firstdefMh}). 

By taking into account this $\rmssm$ Higgs-boson-mass correction, 
we can express
the renormalized vertex functions, in a generic way, as follows:
\bea
\label{eq:firstredef}
{\Gamma_{\,R\,\,h^{0}}^{\,(2)}}&=&-q^2+\mhtree+\Delta \Mh^2+
\Psi_{\rmssm}^{(2)\,\rm {rem}}\,,\nonumber\\
{\Gamma_{\,R\,\,h^{0}}^{\,(3)}}&=&\frac{3g}{2\mz \cw}({\Mh^2}_{\rm{tree}}+\Delta
\Mh^2)+\Psi_{\rmssm}^{(3)\,\rm {rem}}\,,\nonumber\\
{\Gamma_{\,R\,\,h^{0}}^{\,(4)}}&=&\frac{3g^2}{4\mz^2\cw^2}({\Mh^2}_{\rm{tree}}+\Delta
\Mh^2)+\Psi_{\rmssm}^{(4)\,\rm {rem}}\,,
\eea
where all the singular $\Delta_\epsilon$ terms and the logarithmic heavy mass
terms are exclusively contained in $\Delta \Mh^2$. Thus, the apparently
non-decoupling terms are absorbed in the redefinition of the $\Mh$ Higgs-boson
mass. The remainder terms, $\Psi_{\rmssm}^{(2)\,\rm {rem}}$,
$\Psi_{\rmssm}^{(3)\,\rm {rem}}$, and 
$\Psi_{\rmssm}^{(4)\,\rm {rem}}$ in~(\ref{eq:firstredef}) come exclusively  
from the {\it light} particle contributions and are finite. 
 For instance, in the two-point function case we have
\[ 
\Psi_{\rmssm}^{(2)\,\rm {rem}}=\Delta {\Gamma_{\,h^{0}}^{\,(2)}}(q^2)-
\Delta {\Gamma_{\,h^{0}}^{\,(2)}}(\Mh^2) 
\]
with $\Delta{\Gamma_{\,h^{0}}^{\,(2)}}$ given in eqs.~(\ref{eq:lightMSSM}) 
and~(\ref{eq:asympresults2p}). 
For the interpretation of the remainder terms it is crucial to have also the
corresponding one-loop analysis of the $\rsm$ self-interactions, which is done in
the next section. As a result, it can be verified 
that, in the large 
$\ma \gg \mz$ limit and by identifying 
${M^{{\mathrm{tree}}}_{h^0}}^2 \simeq \mz^2 C_{2\beta}^2 \longleftrightarrow
\mSM^{{\mathrm{tree}}\,2}$, 
the remaining terms coincide with the corresponding $\rsm$ ones.

\section{Higgs boson self-couplings in the $\rsm$}
\label{sec:SM}

In the standard $SU(2)_L\times U(1)$ theory, the introduction of 
one scalar field doublet with non-vanishing vacuum expectation value
breaks the gauge symmetry spontaneously to
the electromagnetic subgroup $U(1)$. 
The $\rsm$ Higgs potential
\be
\label{eq:PotSM}
 V(\varphi) = - \mu^2 \, \varphi^{\dagger} \varphi+
 \frac{\lambda}{4} \, (\varphi^{\dagger} \varphi)^2\,,
\ee
contains the complex Higgs doublet
$\varphi$  with hypercharge $Y=1$, and the parameters 
$\lambda$ and $\mu$ related by the vacuum expectation value 
$|\langle \varphi\rangle_0|^2=v^2/2=\mu^2/2 \lambda$.

In order to establish the Higgs mechanism experimentally, the
characteristic self-inter\-act\-ion potential of the $\rsm$
has to be
reconstructed once the Higgs particle has been  discovered. This
task requires the measurement of the
self-couplings of the $\rsm$ Higgs boson. These self-couplings are
uniquely determined by the mass of the Higgs
boson, which is related to the quartic coupling $\lambda$ by $M_{H_{\rsm}}
= \sqrt{2\lambda} \,v$. By introducing the physical Higgs field $H=H_{\rsm}$ in the
neutral component of the doublet, $\varphi^0 = (v+H)/\sqrt{2}$, the tree-level
trilinear and quartic vertices of the Higgs field $H$
can be derived from the potential $V$, yielding
\be
\label{eq:coupSM}
\lambda_{HHH} = \frac{3g \mSMtree}{2\,\mw} 
               = \frac{3 \mSMtree}{v}\,, \,\,\,\quad
\lambda_{HHHH} = \frac{3g^2 \mSMtree}{4\,\mw^2} 
              = \frac{3 \mSMtree}{v^2} \,,
\ee
with $g$ being the ${\rm SU(2)_L}$ gauge coupling. 

We note once again that the $\rmssm$ tree-level 
self-couplings~(\ref{eq:treelevelself}) reach the corresponding $\rsm$
couplings above in the {\it decoupling limit}.
Here we derive the one-loop contributions to the $H_{SM}$ 1PI
Green functions, and in particular those that 
yield the effective triple and quartic self-couplings. 
Concretely, the generic diagrams from the Higgs sector 
contributing to the $n$-point $\rsm$ vertex functions $(n=1,...,4)$
are shown in Fig.~\ref{fig:generic} by choosing $\phi \equiv H_{SM}$ and 
$S\equiv H_{SM}, G^{0},G^{\pm}$. 
The general results for the $n$-point renormalized vertex functions
are summarized by the generic expression~(\ref{eq:notG}). 
The tree-level functions
for the $\rsm$ case ($H\equiv H_{SM}$) and for $n=3,4$
correspond to the  expressions for the $H_{SM}$ Higgs couplings 
already given in~(\ref{eq:coupSM}), and
the one-loop contributions are
summarized in $\Delta \Gamma_{H_{SM}}^{\,(n)}$. 
The computation was done in a general $R_{\xi}$ gauge.
The results for $\Delta \Gamma_{H_{SM}}^{\,(n)}$, in terms  of the 
$2$, $3$ and $4$-point one-loop integrals, are given by
\bea
\label{eq:vertexSM}
&&\hspace*{-2.0cm}{\Delta \Gamma_{H_{\rsm}}^{(1)}}=
\frac{g}{64 \pi^2 \cw}\,\frac{\mSM^2}{\mz}\, \left\{\,
 3 A_{0}(\mSM^2)+A_{0}(\xi \mz^2)+2A_{0}( \xi \mw^2)\,\right\}\,,\nonumber\\
&&\nonumber\\
&&\hspace*{-2.0cm}{\Delta \Gamma_{H_{\rsm}}^{(2)}}=
\frac{g^2}{128 \pi^2 \cw^2}\,\frac{\mSM^2}{\mz^2}\,\left\{\,
3 A_{0}(\mSM^2)+A_{0}(\xi \mz^2)+2A_{0}( \xi \mw^2)\right.\nonumber\\
&+& \left.\mSM^2 \left[\, 
9 B_{0}(q^2,\mSM^2,\mSM^2)+B_{0}(q^2,\xi \mz^2,\xi \mz^2)
+2 B_{0}(q^2,\xi \mw^2,\xi \mw^2)\,\right]\right\},\nonumber\\
&&\nonumber\\
&&\hspace*{-2.0cm}{\Delta \Gamma_{H_{\rsm}}^{(3)}}=
\frac{g^3}{256 \pi^2 \cw^3}\frac{\mSM^4}{\mz^3} \left\{\,\left[\, 
9 B_{0}(q^2,\mSM^2,\mSM^2)+
B_{0}(q^2,\xi \mz^2,\xi \mz^2)\right.\right.\nonumber\\
&&\hspace*{3.0cm}\left.+2 B_{0}(q^2,\xi \mw^2,\xi \mw^2)\,+(q\rightarrow p)+
(q\rightarrow r)\,\right]\nonumber\\
&+&\left.2 \mSM^2 \left[ 
27 C_{0}(q^2,p^2,r^2,\mSM^2,\mSM^2,\mSM^2)+
C_{0}(q^2,p^2,r^2,\xi\mz^2,\xi\mz^2,\xi\mz^2)\right.\right.\nonumber\\
&+&\left.\left.2 C_{0}(q^2,p^2,r^2,\xi\mw^2,\xi\mw^2,\xi\mw^2)\,
\right]\right\},\nonumber\\
&&\nonumber\\
&&\hspace*{-2.0cm}{\Delta \Gamma_{H_{\rsm}}^{(4)}}=
\frac{g^4}{512 \pi^2 \cw^4}\,\frac{\mSM^4}{\mz^4}\,\left\{ 
\left[\, 9B_{0}((q+p)^2,\mSM^2,\mSM^2)
+B_{0}((q+p)^2,\xi \mz^2,\xi \mz^2)\right.\right.\nonumber\\
&+&B_{0}((q+p)^2,\xi \mw^2,\xi \mw^2)
+\left.(p \rightarrow r)+(p \rightarrow t)\right]\nonumber\\
&+&2 \mSM^2  \left[ 
27 C_{0}(q^2,p^2,(q+p)^2,\mSM^2,\mSM^2,\mSM^2)\right.\nonumber\\
&+&C_{0}(q^2,p^2,(q+p)^2,\xi \mz^2,\xi \mz^2,\xi \mz^2)+
2 C_{0}(q^2,p^2,(q+p)^2,\xi \mw^2,\xi \mw^2,\xi \mw^2)\nonumber\\
&&\hspace*{-0.6cm}+\left.(p\rightarrow r)+(p\rightarrow t)+
(q\rightarrow t,p\rightarrow r)+(q\rightarrow p,p\rightarrow r)+
 (q\rightarrow p,p\rightarrow t) \right]\nonumber\\
&+&2 \mSM^4  \left[  
 81D_0(q^2,p^2,r^2,t^2,(q+p)^2,(p+r)^2,
     \mSM^2,\mSM^2,\mSM^2,\mSM^2)\right.\nonumber\\
&+&D_0(q^2,p^2,r^2,t^2,(q+p)^2,(p+r)^2,
     \xi \mz^2,\xi \mz^2,\xi \mz^2,\xi \mz^2)\nonumber\\ 
&+&2D_0(q^2,p^2,r^2,t^2,(q+p)^2,(p+r)^2,
     \xi \mw^2,\xi \mw^2,\xi \mw^2,\xi \mw^2)\nonumber\\
&&+\left.\left.(r\leftrightarrow t)+(p\leftrightarrow r) \right]
\right\}\,.
\eea
These expressions are in general different from the ones obtained in the 
$\rmssm$ [eq.~(\ref{eq:lightMSSM})]. However, 
they acquire the same structure as in the $\rmssm$ in the $\ma \gg \mz$ limit by
identifying the light CP-even Higgs-boson mass with the 
$\rsm$ Higgs mass, that is, $\mSM^2 \longleftrightarrow 
{M^{{\mathrm{tree}}}_{h^0}}^2 \simeq \mz^2 C_{2\beta}^2$.
Consequently, the one-loop {\it light} $\rmssm$
contributions~(\ref{eq:lightMSSM}) converge to the $\rsm$
ones~(\ref{eq:vertexSM}) in the {\it decoupling limit}.
For completeness, we concentrate in the following on the $\rsm$ vertex counterterms
by assuming the on-shell renormalization scheme.

\subsection{On-shell renormalization in the Standard Model}
\label{sec:SMrenor}

The on-shell renormalization scheme for the $\rsm$ has been presented in
previous articles~\cite{jeger,Renor,Denner}, to which we refer 
for details. Here we need only the part for the Higgs sector renormalization.
The counterterms are derived from the 
Higgs potential~(\ref{eq:PotSM}), via multiplicative renormalization,
\bea
&& \varphi \rightarrow  Z_{\varphi}^{1/2} \varphi\,,\nonumber\\
&& \lambda \rightarrow Z_{\lambda} Z_{\varphi}^{-2} \lambda\,,\,\,\,\,
\mu^2 \rightarrow (\mu^2 -\delta \mu^2) Z_{\varphi}^{-1}\,,\nonumber\\
&&v \rightarrow Z_{\varphi}^{1/2}(v -\delta v)\,,\,\,\,\,
\eea
and by expanding  $Z_{i}\rightarrow 1+\delta Z_{i}$. We obtain the following
one-loop counterterms,
\bea
\label{eq:countSM}
{\delta \Gamma_{H_{SM}}^{(1)}}&=& \frac{2 \mz \cw}{g} \mSM^2 \frac{\delta
  t}{t}\,,\,\,\,{\rm{with}}\,\,\, t=\frac{2 \mz \cw\mSM^2}{g}\,,\nonumber\\
{\delta \Gamma_{H_{SM}}^{(2)}}&=& (q^2-M_{H}^2)\,\delta Z_\varphi 
-\delta M_{H}^2\,,\nonumber\\
{\delta \Gamma_{H_{SM}}^{(3)}}&=& -\frac{3 g}{2\mz \cw }\mSM^2
\left(\delta Z_{\lambda}-\frac{\delta v}{v}\right)\,,\nonumber\\
{\delta \Gamma_{H_{SM}}^{(4)}}&=& -\frac{3 g^2}{4\mz^2 \cw^2 }\mSM^2\,
\delta Z_{\lambda}\,.
\eea
with $\delta M_{H}^2$ and $\delta t$ related to the original renormalization constants by
\bea
\label{eq:relacion}
\delta M_{H}^2&=&M_{H}^2\left(-3\frac{\delta v}{v}+\frac{3}{2}\delta Z_{\lambda}-
\delta Z_{\varphi}\right)+\delta \mu^2\,,\nonumber\\
\frac{\delta t}{t}&=&\frac{\delta v}{v}-\frac{\delta \mu^2}{\mSM^2}-
\frac{1}{2} \delta Z_{\lambda}\,.
\eea

In a first step, the counterterms $\delta t/{t}$ and $\delta M_{H}^2$
are determined from two on-shell conditions in the Higgs sector:\\
- vanishing renormalized tadpole diagram: 
\be
\label{eq:cond1SM}
{\Delta \Gamma_{R\,H_{SM}}^{(1)}}={\Delta \Gamma_{H_{SM}}^{(1)}}
+{\delta \Gamma_{H_{SM}}^{(1)}}=0\,,
\ee
- the pole of the renormalized Higgs propagator lies at $\mSM^2$, which implies
\be
\label{eq:cond2SM}
{\Delta \Gamma_{R\,H_{SM}}^{(2)}}(\mSM^2)={\Delta \Gamma_{H_{SM}}^{(2)}}(\mSM^2)+
 {\delta \Gamma_{H_{SM}}^{(2)}}(\mSM^2)=0.
\ee
Solving these equations, we obtain
\bea
\label{eq:resultscountSM}
\hspace*{-0.5cm}\frac{\delta t}{t}&=&- 
\frac{g^2}{128 \pi^2 \cw^2}\,\frac{1}{\mz^2}\, \left\{ 3\, A_{0}(\mSM^2)+
A_{0}( \xi \mz^2)+2\,A_{0}( \xi \mw^2)\right\}\,,\nonumber\\
\nonumber\\
\hspace*{-0.5cm}\delta M_{H}^2&=& 
\frac{g^2}{128 \pi^2 \cw^2}\,\frac{\mSM^2}{\mz^2}\,\left\{
\,\left[3\, A_{0}(\mSM^2)+A_{0}( \xi \mz^2)+
2\,A_{0}( \xi \mw^2)\right]\right.\nonumber\\
&+&\mSM^2\, \left[9B_{0}(\mSM^2,\mSM^2,\mSM^2)+
B_{0}(\mSM^2,\xi \mz^2,\xi \mz^2)\right.\nonumber\\
&&\left.\left.+2B_{0}(\mSM^2,\xi \mw^2,\xi \mw^2)\right]\,\right\}.
\eea

Next, we need a condition to fix the field-renormalization constant
$\delta Z_{\varphi}$. The conventional on-shell condition would
be to require unity residue for the physical Higgs-boson propagator, 
yielding
\bea
\label{eq:deltaZphi}
\hspace*{-0.5cm}\delta Z_{\varphi}&=&\frac{g^2}{128 \pi^2 \cw^2}\,
\frac{\mSM^2}{\mz^2}\,\left\{
9B^{'}_{0}(\mSM^2,\mSM^2,\mSM^2)+
B^{'}_{0}(\mSM^2,\xi \mz^2,\xi \mz^2)\right.\nonumber\\
&&\left.+2B^{'}_{0}(\mSM^2,\xi \mw^2,\xi \mw^2)\,\right\},
\eea
which is different from zero.
For our purpose of comparing the SM and MSSM vertex functions, however,
this appears to be inconvenient because in the large $M_A$ limit of the MSSM
the Higgs field-renormalization constants vanish, as discussed in section
\ref{sec:renorMSSM}, and thus the external lines would carry different
normalizations in the two models.
It is therefore more natural to adopt for the SM a condition that leads to
the same normalization and to require 
\be 
\delta Z_{\varphi} = 0 \, ,
\ee
instead of eq.~(\ref{eq:deltaZphi}). This is possible because
$\delta Z_{\varphi}$ is a UV-finite quantity.
With this condition we can compare the two models 
directly on the basis of the irreducible 
renormalized vertex functions.

Unlike the previous ones, the $\delta v$ renormalization constant is
determined from the gauge sector.
We have checked explicitly that the result for $\delta v$ in the SM
corresponds to the result for $\delta v$ in the $\rmssm$ 
whenever the $\ma\gg\mz$ limit is
considered and by identifying $\Mh\leftrightarrow \mSM$.
Thus, the expression for $\delta v/v$ in the $\rsm$ can be obtained 
from~(\ref{eq:vyGMSSM}) by simply replacing $\Mh$ by $\mSM$.

Finally, $\delta Z_{\lambda}$ is determined  
with the help of the relation~(\ref{eq:relacion}):
\bea
\label{eq:deltaZL}\hspace*{-0.8cm}
\delta Z_{\lambda}&=& \frac{g^2}{128 \cw^2 \pi^2}\frac{1}{\mz^2}
\left\{2A_{0}(\mSM^2)+2A_{0}(\xi \mz^2)+
\frac{4}{\sw^2}(1-\cw^2-4\cw^4+4\cw^6)A_{0}( \xi \mw^2)\right.\nonumber\\
&-&\frac{8}{3}\cw^2\sw^2\mz^2B_{0}(0,\xi \mw^2,\xi \mw^2)\nonumber\\
&+&\mSM^2\left[9B_{0}(\mSM^2,\mSM^2,\mSM^2)+2B_{0}(\mSM^2,\xi \mw^2,\xi
  \mw^2)\right.\nonumber\\
&+&\left.B_{0}(\mSM^2,\xi \mz^2,\xi\mz^2)\right]
-8\frac{\cw^2}{\sw^2}B_{22}(\mz^2, \mSM^2,\xi \mz^2)\nonumber\\
&-&\frac{8}{\sw^2}(1-2\cw^2)\left[
B_{22}(\mw^2, \mSM^2,\xi \mw^2)+B_{22}(\mw^2, \xi \mz^2,\xi
\mw^2)\right.\nonumber\\
&+& \left.
\left.\cw^2(1-2\cw^2)B_{22}(\mz^2, \xi \mw^2,\xi \mw^2)\right] \right\} \,.
\eea

The corresponding vertex counterterms follow 
immediately via substitution of the 
$\rsm$ renormalization constants in~(\ref{eq:countSM}). 
The $\rsm$
renormalized vertex functions are easily obtained by adding the one-loop
contributions~(\ref{eq:vertexSM})
and the corresponding counterterms~(\ref{eq:countSM})-(\ref{eq:deltaZL}).
Remember that the renormalized one-point $\rsm$ 
vanishes in the present on-shell renormalization scheme. In addition, 
the renormalized two-point $\rsm$ function also vanishes at the physical mass 
$M_{H_{\rsm}}^2$,
but not at general $q^2$. According to the fact that the 
renormalized Higgs boson self-energy evaluated at the physical mass-squared defines the
Higgs-boson mass correction, we obtain trivially that 
$\Delta \mSM^2=0$ (this is nothing else than the on-shell mass condition, which
implies $\mSM^2=\mSM^{{\rm tree}\,2}\,$).

Together with
the tree-level $\rsm$ Higgs self-interactions~(\ref{eq:coupSM}), 
the renormalized trilinear and quartic $H_{\rsm}$
vertex functions at the one-loop level can be written as
\bea
\label{eq:firstredefSM}
{\Gamma_{\,R\,\,H_{SM}}^{\,(3)}}&=&\frac{3g}{2\mz \cw}\,{\mSM^2}+
\Psi_{SM}^{(3)\,\rm{rem}}\,,\nonumber\\
{\Gamma_{\,R\,\,H_{SM}}^{\,(4)}}&=&\frac{3g^2}{4\mz^2\cw^2}{\mSM^2}+
\Psi_{SM}^{(4)\,\rm{rem}}\,,
\eea
where $\Psi_{SM}^{(3)\,\rm{rem}}$ and $\Psi_{SM}^{(4)\rm{rem}}$ are  
UV-finite functions depending on the external momenta.
Remember that similar finite terms 
were obtained from the {\it light} contributions  
in the $\rmssm$ case,  
summarized in $\Psi_{\rmssm}^{(3)\,\rm {rem}}$ and 
$\Psi_{\rmssm}^{(4)\,\rm {rem}}$  in eq.~(\ref{eq:firstredef}). 
For arbitrary $\ma$ values, these finite contributions are different in both
models. However, for large $\ma$  and by identifying 
${M^{{\mathrm{tree}}}_{h^0}}^2 \simeq \mz^2 C_{2\beta}^2 \longleftrightarrow
\mSM^2$, we obtain that  
$\Psi_{MSSM}^{(n)\rm{rem}}\longrightarrow \Psi_{SM}^{(n)\rm{rem}}$ $(n=3,4)$. 
Thus, these
contributions coincide in the $\ma \gg \mz$ limit and do not lead to differences between
the two models.

\section{Discussion and conclusions} 
\label{sec:Matching}

In order to study whether the $\rsm$ Higgs
sector can be considered as the low-energy effective theory of the $\rmssm$ Higgs
sector in the $\ma \gg \mz$ limit we have compared the predictions in the two
theories of the renormalized $n$-point 1PI Green functions 
for  $h^{0}$ and $H_{\rsm}$, respectively, at the
one-loop level, and for $n=1,...,4$.  We have
examined in full detail the veracity of the equality among these functions by 
comparing them at low-energy scales $p^2\ll\ma^2$ and by 
choosing a particular renormalization scheme, the on-shell scheme. 
This matching~\cite{Matching} between the two theories, {\it via} 
renormalized vertex functions, can be summarized by
\be
\label{eq:condmatching}
{\Gamma_{\,R\,\,h^{0}}^{\,(n)}}^{\rmssm}(p)=
{\Gamma_{\,R\,\,H_{\rsm}}^{\,(n)}}^{\rsm}(p)\,\,,\,\,\,\,\,\,\,\,p\ll\ma
\hspace*{1.5cm} (n=1,...,4)
\ee
where the left-hand side must be understood
as the $\rmssm$ functions in the $\ma \gg \mz$ limit. 

It is worth emphasizing now some important points regarding this comparison 
of the vertex functions of the two theories.
First, as stated in sections~\ref{sec:Spectrum} and~\ref{sec:SM},
the tree-level self-couplings in both models
[see eq.~(\ref{eq:treelevelself}) and~(\ref{eq:coupSM})] lead to equal results
in the $\rsm$ and in the $\rmssm$ vertex functions in the 
{\it decoupling limit}. 
This implies that the tree-level contributions
can be dropped from both sides of the matching 
conditions~(\ref{eq:condmatching}).
Second, as explained in section~\ref{sec:Spectrum}, 
the subset of diagrams that have any number of gauge bosons 
in the loops gives the same contributions in the $\rsm$ and in the 
$\rmssm$ (in the $\ma \gg \mz$ limit) and, therefore, these can also be 
dropped from both sides of the matching 
conditions~(\ref{eq:condmatching}). In fact, these kinds of contribution 
have not been considered explicitly in the present computation. 
Third, diagrams involving just Goldstone bosons and the lightest Higgs boson
in the loops 
do contribute with non-vanishing corrections which, in principle,
are not the same in both models. However,
we have demonstrated that the one-loop contributions, given
in~(\ref{eq:lightMSSM}) and~(\ref{eq:vertexSM}) in the $\rmssm$ and the $\rsm$,
respectively, coincide in the $\ma \gg \mz$ limit. Therefore, 
they do not contribute either to the differences between the two models in 
the matching conditions~(\ref{eq:condmatching}). In contrast, we found some
{\it light} contributions to the vertex counterterms 
[see eqs.~(\ref{eq:redeflight}) and~(\ref{eq:countSM})-(\ref{eq:deltaZL})] that are different in both models.
These differences in the {\it light} sector come from the fact that, whereas the
$\delta v/v$ contributions are the same in the $\rsm$ and in the
$\rmssm$ in the $\ma \gg \mz$ limit, the other renormalization constants,
that is, $\delta Z_{\lambda}$ in the $\rsm$ and $\delta G^2$ in the $\rmssm$,
do not coincide. 
The mass counterterms for the $h^{0}$ and $H_\rsm$ fields 
do not coincide either.
Thus, what we understand by {\it light} contributions in this
work are also important in the differences between the renormalized vertex 
functions in both theories.

Overall, we can say that the
differences between the one-loop renormalized vertex functions of the
two theories in the {\it decoupling limit} come, on the one hand, from the
one-loop diagrams including at least one heavy $\rmssm$ Higgs particle and, on
the other hand, from the vertex counterterms.
Concretely, eqs.~(\ref{eq:asympresults1p})-~(\ref{eq:asympresults4p}) give the
  differences between the one-loop unrenormalized vertex functions of
  the two theories. Consequently, they cannot be dropped in the 
conditions~(\ref{eq:condmatching}).
Moreover, these different contributions have a finite piece that depends
logarithmically and quadratically on the heavy Higgs-boson mass $\ma$ and a 
divergent piece in $D=4$, and both pieces 
summarize the potential non-decoupling effects of the heavy Higgs-boson sector
of the $\rmssm$.
It is essential, however, that these heavy Higgs particle effects can
be absorbed into redefinitions of the low energy parameters, thus not
providing any physically observable effect~\cite{App-Cara}. 
As we have seen,
the counterterms in the $\rsm$ and in the $\rmssm$
are different in both models and therefore
they also contribute to the differences between the two models in
the matching conditions~(\ref{eq:condmatching}).

Putting all results together and comparing eqs.~(\ref{eq:firstredef}) 
and~(\ref{eq:firstredefSM}), the differences found in the 
unrenormalized vertex functions are exactly
compensated by the $\Delta \Mh^2$ contribution, and the final results for the
renormalized $2,3$- and $4$-point functions coincide in the two models in the
large $\ma \gg \mz$ limit, as required by the matching 
conditions~(\ref{eq:condmatching}). In other words, all the
potential non-decoupling effects from the heavy Higgs modes can be absorbed into 
the redefinition of the lightest Higgs boson mass
$\Mh$ [see eqs.~(\ref{eq:firstdefMh}) and~(\ref{eq:firstredef})] 
and therefore decoupling of the heavy $\rmssm$ Higgs particles occurs.
We notice that, for arbitrary $\ma$ value, there are other finite terms 
in the renormalized $\rmssm$  $n$-point
functions, summarized by the remainder parts $\Psi_{\rmssm}^{(3)\,\rm {rem}}$ 
and $\Psi_{\rmssm}^{(4)\,\rm {rem}}$ of 
eq.~(\ref{eq:firstredef}). However, we have shown 
that in the $\rsm$ similar contributions appear in the
renormalized $H_{\rsm}$ vertex functions, summarized by the remainder parts 
$\Psi_{\rsm}^{(3)\,\rm {rem}}$ and $\Psi_{\rsm}^{(4)\,\rm {rem}}$
of eq.~(\ref{eq:firstredefSM}), which coincide with the corresponding
$\rmssm$ terms in the large $\ma$ limit. 
Therefore, these contributions drop out as well in the 
matching conditions~(\ref{eq:condmatching}).

In conclusion, we have demonstrated that all the apparent
non-decoupling one-loop effects from the heavy $\rmssm$ Higgs bosons 
are absorbed in the $\rmssm$ Higgs-boson mass $\Mh$, and the remaining 
contributions are suppressed by inverse powers of $\ma$ and
therefore vanish in the large $\ma$ limit. Thus,  
the $h^{0}$ self-interactions converge to the $H_{\rsm}$ 
self-interactions at the one-loop level and in the $\ma \gg \mz$ limit, 
and the $\rmssm$ $h^{0}$ self-couplings thereby acquire the same structure 
as the couplings of the $\rsm$ Higgs boson whenever one identifies 
$\Mh\leftrightarrow\mSM$. 
Equivalently, we showed that  the heavy $\rmssm$
Higgs sector decouples from low
energy, at the electroweak scale, and leaves behind the $\rsm$ 
Higgs sector in the Higgs self-interactions also. 
Consequently, we would need extremely high-precision experiments for the 
experimental verification of the SUSY nature of the Higgs boson self-interactions.

\vspace{0.3cm}
\section*{Acknowledgments}
 
\noindent The work of S.P. was
supported by the \textit{Fundaci{\'o}n Ram{\'o}n Areces}.
Support by the European Union under HPRN-CT-2000-00149
and by the Spanish Ministerio de Ciencia y Tecnolog{\'{\i}}a under CICYT projects 
FPA 2000-0980, FPA 2000-0956 and PB98-0782 is gratefully acknow\-ledged.

\section*{Appendix A.}
\setcounter{equation}{0}
\renewcommand{\theequation}{A.\arabic{equation}}

In this appendix we display, first, the formulas for the 
one-loop contributions to the $H^{0}$ tadpole diagrams and the $A^{0}$ boson
self-energies that are required for on-shell renormalization.
Next we present the ${\cal O}\left({\mz^2}/{\ma^2}\right)$ 
contributions to the renormalization constants $\delta Z_{H_{1,2}}$, which 
are relevant in order to impose 
the $A^{0}$-boson on-shell condition. Finally, results for the 
$\delta m_1\,, \delta m_2,$ and $\delta m_{12}$ mass counterterms are given.
Here we follow the notation introduced 
throughout this article for {\it light}, {\it mixed} and {\it heavy}
contributions, as explained in~(\ref{eq:notacion}).

\vspace*{0.5cm}
$\bullet$ $H^{0}$ tadpole and $A^{0}$ boson self-energies:
\bea
\label{eq:H0}
{\Delta \Gamma_{H^{0}}^{(1)}}^{\rm{light}}&=&
\frac{g \mz}{64 \pi^2 \cw}\,C_{2\beta}\, S_{2\beta}\, \left\{
 3 A_{0}(\Mh^2)+A_{0}(\xi \mz^2)+2 A_{0}(\xi \mw^2)\right\}\,,\nonumber\\
\nonumber\\
{\Delta \Gamma_{H^{0}}^{(1)}}^{\rm{heavy}}&=&
-\frac{g \mz}{32 \pi^2 \cw}\,\,C_{2\beta}\, S_{2\beta}\, \left\{ 3\ma^2 \,\left(
\Delta_\epsilon +1-\log\frac{\ma^2}{\mu_{0}^2}\right)\right.\nonumber\\
&&-\frac{1}{2}\mz^2\left. \left[ (6-12 C_{2\beta}^2+4\cw^2)+(3-9C_{2\beta}^2+2 \cw^2)
\left(\Delta_\epsilon-\log\frac{\ma^2}{\mu_{0}^2}\right)\right]\right\},\nonumber\\
\eea
\bea
\label{eq:A0}
{\Delta \Gamma_{A^{0}}^{(2)}}^{\rm{light}}(\ma^2)&=&
-\frac{g^2}{128 \pi^2 \cw^2}\,\left\{C_{2\beta}^2\,A_{0}(\Mh^2)
- (2-3C_{2\beta}^2) A_{0}(\xi \mz^2)\right.\nonumber\\
&&+\left.2 (C_{2\beta}^2-2\cw^2) A_{0}(\xi \mw^2)-
2C_{2\beta}^2\, S_{2\beta}^2 \mz^2\,B_{0}(\ma^2,\Mh^2,\xi\mz^2)\right\},\nonumber\\
\nonumber\\
{\Delta \Gamma_{A^{0}}^{(2)}}^{\rm{mixed}}(\ma^2)&=&
\frac{g^2}{64 \pi^2 \cw^2}\mz^2 (1+2\cw^4+2C_{2\beta}^4-2C_{2\beta}^2)
\left(\Delta_\epsilon+2-\log\frac{\ma^2}{\mu_{0}^2}\right),\nonumber\\
{\Delta \Gamma_{A^{0}}^{(2)}}^{\rm{heavy}}(\ma^2)&=&
\frac{g^2}{64 \pi^2 \cw^2}\,C_{2\beta}^2\left\{
\mz^2 (\cw^2+\frac{1}{2}S_{2\beta}^2)
\left(\Delta_\epsilon-\log\frac{\ma^2}{\mu_{0}^2}\right)\right.\nonumber\\
&&\left.+3\ma^2
\left(\Delta_\epsilon+1-\log\frac{\ma^2}{\mu_{0}^2}\right)
+\mz^2 (1-C_{2\beta}^2)\left(1-\frac{\pi}{\sqrt{3}}\right)\right\}\,.\nonumber\\
\eea

\vspace*{0.3cm}
$\bullet$ $\delta Z_{H_{1}}$ and $\delta Z_{H_{2}}$ counterterms:
\bea
\label{eq:ZH12}
\delta Z_{H_{1}}^{\rm{light}}&=&-\frac{g^2}{64 \pi^2 \cw^2}\,
C_{2\beta}\, S_{2\beta}\left[\mz^2\, C_{2\beta}\,S_{2\beta}\,
B_{0}^{'}(\ma^2,\Mh^2,\xi\mz^2)\right.\nonumber\\
&&\hspace*{3.2cm}+\left.
\cot\beta\,(B_{0}+2B_{1})(\ma^2,\Mh^2,\xi\mz^2)\,\right]\,,\nonumber\\
\delta Z_{H_{1}}^{\rm{mixed}}&=&\frac{g^2}{64 \pi^2 \cw^2}\,
\frac{\mz^2}{\ma^2}\,\left\{\left(1+2\cw^4+2 C_{2\beta}^4
-2C_{2\beta}^2\right)-\cot\beta\, C_{2\beta}\,S_{2\beta}\right\}\,,\nonumber\\
\delta Z_{H_{1}}^{\rm{heavy}}&=&\frac{g^2}{64 \pi^2 \cw^2}\,
\frac{\mz^2}{\ma^2}\,\left\{C_{2\beta}^2\, S_{2\beta}^2
\left(1-\frac{2\pi}{3\sqrt{3}}\right)+\cot\beta \,C_{2\beta}\, S_{2\beta}^3
\left(2-\frac{\pi}{\sqrt{3}}\right)\right\}\,,\nonumber\\
\nonumber\\
\delta Z_{H_{2}}^{\rm{light}}&=&-\frac{g^2}{64 \pi^2 \cw^2}\,
C_{2\beta}\, S_{2\beta}\left[\mz^2\, C_{2\beta}\,S_{2\beta}\,
B_{0}^{'}(\ma^2,\Mh^2,\xi\mz^2)\right.\nonumber\\
&&\hspace*{3.2cm}-\left.
\tan\beta\,(B_{0}+2B_{1})(\ma^2,\Mh^2,\xi\mz^2)\,\right]\,,\nonumber\\
\delta Z_{H_{2}}^{\rm{mixed}}&=&\frac{g^2}{64 \pi^2 \cw^2}\,
\frac{\mz^2}{\ma^2}\,\left\{\left(1+2\cw^4+2 C_{2\beta}^4
-2C_{2\beta}^2\right)+\tan\beta\, C_{2\beta}\,S_{2\beta}\right\}\,,\nonumber\\
\delta Z_{H_{2}}^{\rm{heavy}}&=&\frac{g^2}{64 \pi^2 \cw^2}\,
\frac{\mz^2}{\ma^2}\,\left\{C_{2\beta}^2\, S_{2\beta}^2
\left(1-\frac{2\pi}{3\sqrt{3}}\right)-\tan\beta \,C_{2\beta}\, S_{2\beta}^3
\left(2-\frac{\pi}{\sqrt{3}}\right)\right\}\,.
\eea

\vspace*{0.3cm}
$\bullet$ $\delta m_1\,, \delta m_2$ and $\delta m_{12}$ counterterms:

\bea
\label{eq:mslight}
\hspace*{-0.6cm}\delta m_{1}^{2\,\rm{light}}&=&
\frac{g^2}{512 \pi^2 \cw^2}\,\left\{16 C_{2\beta}A_0(\Mh^2)
-4C_{2\beta}B_{22}(\mz^2,\Mh^2,\xi \mz^2) (5-2C_{2\beta}+C_{4\beta})\right.\nonumber\\ 
&-&2(1-2\cw^2)^2B_{22}(\mz^2,\xi \mw^2,\xi \mw^2)
(-2+11C_{2\beta}-2C_{4\beta}+C_{6\beta})\nonumber\\
&+&2A_0(\xi \mz^2)(-2+8\cw^2+(7-8\cw^2)C_{2\beta}-2C_{4\beta}+C_{6\beta})\nonumber\\
&+&A_0(\xi \mw^2)(2+8\cw^2-8\cw^4+(17-44\cw^2+44\cw^4)C_{2\beta}
-2(3-4\cw^2+4\cw^4)C_{4\beta}\nonumber\\
&+&3C_{6\beta}-4\cw^2C_{6\beta}+4\cw^4C_{6\beta})
+64\mz^2 C_{\beta}^2 C_{2\beta}^2 S_{\beta}^4 B_0(\ma^2,\Mh^2,\mz^2)\nonumber\\
&-&\left.64\ma^2 \mz^2 C_{2\beta}^2 C_{\beta}^2 S_{\beta}^4
B'_0(\ma^2,\Mh^2,\xi\mz^2)\,\right\}, \nonumber\\
\nonumber\\
\hspace*{-0.6cm}\delta m_{2}^{2\,\rm{light}}&=&
\frac{g^2}{256 \pi^2 \cw^2}\,\left\{
-8C_{2\beta}A_0(\Mh^2)+(2+11 C_{2\beta}+2C_{4\beta}+C_{6\beta})
B_{22}(\mz^2,\Mh^2,\xi \mz^2)\right.\nonumber\\
&+&(1-2\cw^2)^2(2+11 C_{2\beta}+2C_{4\beta}+C_{6\beta})
B_{22}(\mz^2,\xi \mw^2,\xi \mw^2)\nonumber\\
&-&\mz^2 C_{\beta}^2 (-1+C_{8\beta})B_0(\ma^2,\Mh^2,\mz^2)\nonumber\\
&+&\ma^2\mz^2 C_{\beta}^2 (C_{8\beta}-1)B'_0(\ma^2,\Mh^2,\xi\mz^2)\nonumber\\
&+&4A_0(\xi \mz^2)(-C_{\beta}^2(2-4\cw^2+C_{4\beta})+S_{\beta}^2)
-2A_0(\xi \mw^2)(C_{\beta}^2(3(1-2\cw^2)^2\nonumber\\
&+&\left.(3-4\cw^2+4\cw^4)C_{4\beta})
-4(1-2\cw^2+2\cw^4)S_{\beta}^2)\,\right\}, \nonumber\\
\nonumber\\
\hspace*{-0.6cm}\delta m_{12}^{2\,\rm{light}}&=&
-\frac{g^2}{128 \pi^2 \cw^2}C_{\beta}S_{\beta}\,\left\{
4 C_{2\beta}^2B_{22}(\mz^2,\Mh^2,\xi \mz^2)\right.\nonumber\\
&+&4(1-2\cw^2)^2 C_{2\beta}^2 B_{22}(\mz^2,\xi \mw^2,\xi \mw^2)
-2A_0(\xi \mz^2)(1-4\cw^2+C_{4\beta})\nonumber\\
&-&A_0(\xi \mw^2)(-1-4\cw^2+4\cw^4+(3-4\cw^2+4\cw^4)C_{4\beta})\nonumber\\
&+&\left.\mz^2 S_{4\beta}^2 B_0(\ma^2,\Mh^2,\mz^2)-
 \ma^2\mz^2 S_{4\beta}^2 B'_0(\ma^2,\Mh^2,\xi\mz^2)\,\right\},
\eea

\bea
\label{eq:m12mixed}
\delta m_{1}^{2\,\rm{mixed}}&=&\frac{g^2}{32 \pi^2 \cw^2}\,
\mz^2 \,\sin^2\beta\,\left(1+2\cw^4+2 C_{2\beta}^4
-2C_{2\beta}^2\right)\left(\Delta_\epsilon+3-
\log\frac{\ma^2}{\mu_{0}^2}\right)\,,\nonumber\\
\delta m_{2}^{2\,\rm{mixed}}&=&\frac{g^2}{32 \pi^2 \cw^2}\,
\mz^2 \,\cos^2\beta\,\left(1+2\cw^4+2 C_{2\beta}^4
-2C_{2\beta}^2\right)\left(\Delta_\epsilon+3-
\log\frac{\ma^2}{\mu_{0}^2}\right)\,,\nonumber\\
\delta m_{12}^{2\,\rm{mixed}}&=&-\frac{g^2}{64 \pi^2 \cw^2}\,
\mz^2 \,S_{2\beta}\,\left(1+2\cw^4+2 C_{2\beta}^4
-2C_{2\beta}^2\right)\left(\Delta_\epsilon+3-
\log\frac{\ma^2}{\mu_{0}^2}\right)\,.\nonumber\\
\eea
\bea
\label{eq:msheavy}
\delta m_{1}^{2\,\rm{heavy}}&=&\frac{g^2}{2304 \pi^2\cw^2}\,
\left\{\frac{9}{2}\ma^2(10+8\cw^2-C_{2\beta}(29-8\cw^2)
+6C_{4\beta}-3C_{6\beta})\right.\nonumber\\
&&\times
\left(\Delta_\epsilon\nonumber+1-\log\frac{\ma^2}{\mu_{0}^2}\right)\nonumber\\
&+&C_{2\beta}S_{2\beta}^2\,\mz^2(9+10\sqrt{3}\pi+(-2C_{2\beta}+C_{4\beta})
(-63+10\sqrt{3}\pi))\nonumber\\
&+&\frac{3}{16}
\mz^2\left[126-48\cw^2+288\cw^4-2(59-4\cw^2+40\cw^4)C_{2\beta}\right.\nonumber\\
&+&48\cw^2(-1+2\cw^2)C_{4\beta}-3(19-8\cw^2+16\cw^4)C_{6\beta}\nonumber\\
&-&\left.\left.30C_{8\beta}+15C_{10\beta}\right]
\left(\Delta_\epsilon\nonumber-\log\frac{\ma^2}{\mu_{0}^2}\right)\right\}\,,\nonumber\\
\delta m_{2}^{2\,\rm{heavy}}&=&\frac{g^2}{2304 \pi^2\cw^2}\,
\left\{\frac{9}{2}\ma^2(10+8\cw^2-C_{2\beta}(8\cw^2-29)
+6C_{4\beta}+3C_{6\beta})\right.\nonumber\\
&&\times
\left(\Delta_\epsilon\nonumber+1-\log\frac{\ma^2}{\mu_{0}^2}\right)\nonumber\\
&-&C_{2\beta}S_{2\beta}^2\,\mz^2(9+10\sqrt{3}\pi+
(-2C_{2\beta}-C_{4\beta})(63-10\sqrt{3}\pi))\nonumber\\
&+&\frac{3}{16}
\mz^2\left[126-48\cw^2+288\cw^4+2(59-4\cw^2+40\cw^4)C_{2\beta}\right.\nonumber\\
&+&48\cw^2(-1+2\cw^2)C_{4\beta}+3(19-8\cw^2+16\cw^4)C_{6\beta}\nonumber\\
&-&\left.\left.30C_{8\beta}-15C_{10\beta}\right]
\left(\Delta_\epsilon\nonumber-\log\frac{\ma^2}{\mu_{0}^2}\right)\right\}\,,\nonumber\\
\delta m_{12}^{2\,\rm{heavy}}&=&\frac{g^2}{2304 \pi^2
  \cw^2}\,S_{2\beta}\,\left\{18\ma^2(1-3C_{2\beta}^2+2\cw^2)
\left(\Delta_\epsilon\nonumber+1-\log\frac{\ma^2}{\mu_{0}^2}\right)\right.\nonumber\\
&-&2C_{2\beta}^2\mz^2\left[45-10\sqrt{3}\pi-C_{2\beta}^2(63-10\sqrt{3}\pi)
\right]\nonumber\\
&+&\left.
9\mz^2\left[2+5C_{2\beta}^4+4\cw^4-C_{2\beta}^2(5-2\cw^2+4\cw^4)\right]
\left(\Delta_\epsilon\nonumber-\log\frac{\ma^2}{\mu_{0}^2}\right)\right\}\,.\nonumber\\
\eea
Thereby, $S_{\beta}\equiv \sin \beta\,,\,C_{\beta}\equiv \cos \beta\,,\,
C_{4\beta}\equiv \cos 4 \beta\,,\, C_{6\beta}\equiv \cos 6 \beta\,,
\,C_{8\beta}\equiv \cos 8 \beta\,,$ etc., are used  for
abbreviations.

\begingroup\raggedright

\end{document}